\RequirePackage{ifpdf}
\documentclass[12pt]{article}
\usepackage[T1]{fontenc}
\usepackage{xcolor}
\usepackage{amsmath, amsfonts, amsthm, amssymb, graphicx, wasysym}
\usepackage{stackengine}
\usepackage{xspace}
\usepackage{epsfig}
\usepackage{psfrag}
\usepackage{geometry}
\usepackage{colonequals}
\usepackage{hyperref}
\usepackage{tikz}
\usepackage[capitalise]{cleveref}
\hypersetup{colorlinks=true,citecolor=black,linkcolor=black,urlcolor=black}
\usepackage{subfig}

\pdfoutput=1

\newcommand{\op}{\ensuremath{\mathcal{O}}\xspace}

\newcommand{\vev}[1]{\ensuremath{\langle #1 \rangle}\xspace}

\renewcommand{\Re}{\text{Re}}

\newcommand{\hf}{\frac{1}{2}}

\let\a=\alpha \let\b=\beta \let\g=\gamma \let\d=\delta \let\e=\epsilon
    \let\k=\kappa
    \let\r=\rho
    \let\c=\chi 
  \let\D=\Delta  
    
    \let\G=\Gamma
\newcommand{\hb}{{\bar h}}
\newcommand{\zb}{{\bar z}}

\newcommand{\ab}{{\bar \alpha}}
\newcommand{\bb}{{\bar \beta}}
\newcommand{\fh}{\hat f}
\newcommand{\ft}{\check f}
\newcommand{\ktil}{{\tilde k}}
\newcommand{\fd}{{\diam{f}}}

\newcommand{\bxx}[1]{\begin{#1}}
	\newcommand{\be}{\bxx{equation}}
	\newcommand{\ee}{\end{equation}}

\def\ea{\end{array}}

\numberwithin{equation}{section}

\usepackage{scalerel}
\usepackage{stackengine,wasysym}

\newcommand\reallywidetilde[1]{\ThisStyle{%
  \setbox0=\hbox{$\SavedStyle#1$}%
  \stackengine{-.1\LMpt}{$\SavedStyle#1$}{%
    \stretchto{\scaleto{\SavedStyle\mkern.2mu\AC}{.5150\wd0}}{.6\ht0}%
  }{O}{c}{F}{T}{S}%
}}

\newcommand{\diam}[1]{\overset{\diamond}{#1}}

\title{\textbf{Applications of Alpha Space}}
\date{}
\author{Daniel Rutter$^\a$, Balt C. van Rees$^{\a,\ab}$ \\\\
\textsl{\small $^\a$Centre for Particle Theory, Department of Mathematical Sciences}\\
\textsl{\small Durham University, DH1 3LE, UK}\\
\textsl{\small $^\ab$CPHT, CNRS, \'Ecole Polytechnique, Institut Polytechnique de Paris}\\
\textsl{\small Route de Saclay, 91128 Palaiseau, France}\\
}
\begin{document}
\maketitle
\begin{abstract}
\thispagestyle{empty}
We extend the definition of `alpha space' as introduced in \cite{Hogervorst:2017sfd} to two spacetime dimensions. We discuss how this can be used to find conformal block decompositions of known functions and how to easily recover several lightcone bootstrap results. In the second part of the paper we establish a connection between alpha space and the Lorentzian inversion formula of \cite{Caron-Huot:2017vep}.
\end{abstract}
\newpage
\tableofcontents
\newpage
\section{Introduction}
The authors of \cite{Hogervorst:2017sfd} introduced `alpha space', a special case of a Jacobi transform \cite{flensted1973convolution,flensted1979jacobi,TK1}, as a means to study correlation functions $g(z)$ of one-dimensional conformally invariant systems. In this paper we discuss a natural two-dimensional version of this construction and discuss how it can be used to analyze correlation functions $g(z,\zb)$ that have a convergent decomposition in two-dimensional (global) conformal blocks. For simplicity we will restrict ourselves to correlation functions of identical operators, but we believe that our results can easily be extended to the more general case.

In the next section we recall the one-dimensional definitions of \cite{Hogervorst:2017sfd} and propose the two-dimensional extension. Alpha space appears to be particularly useful for finding the OPE coefficients of the conformal block decomposition in closed form, as we will demonstrate by proving some exact identities. In section 3 we show that we can easily derive some of the well-known `lightcone bootstrap' results for conformal correlators with a twist gap in the $t$-channel. We believe that alpha space could also be helpful in the analysis of subleading terms related to the exponentation of anomalous dimensions and the summation of entire Regge trajectories, which we illustrate by working out the leading order effects of exponentiation.

In some sense our alpha space transform is inherently Lorentzian because it is an integral transform over the square where $0 < z, \zb < 1$. What is not built in, therefore, is the single-valuedness of correlation functions in Euclidean signature, \emph{i.e.}, the requirement that the spins of the conformal blocks must be integers. This makes it manifestly different from the density that is defined using harmonic analysis in \cite{Dobrev:1977qv} and which is now known to admit an analytic continuation in spin through the `Lorentzian inversion formula' of \cite{Caron-Huot:2017vep}. However we show in section 4 that it is nevertheless possible to establish a connection between the densities. To do so we will define a slightly \emph{different} integral transform of conformal correlators which only works if the spins are integers. We will show that our alpha space results can be used to determine the meromorphicity properties of this density, but also that some simple contour pulling directly relates it to the Lorentzian inversion formula of \cite{Caron-Huot:2017vep}. One may view this as providing a (heuristic) derivation that the Lorentzian inversion formula indeed reproduces the conformal OPE coefficients for integer and sufficiently large spins - but now starting from alpha space rather than from the Euclidean inversion formula.

One important question not addressed here is the existence of an alpha space transform in more than two dimensions. We hope to address this in the near future. See \cite{Dansthesis} for a preview and also \cite{Isachenkov:2016gim,Isachenkov:2017qgn} for related developments.

\section{Alpha space in two dimensions}
Consider first an integrable function $f(z)$ defined for $0 < z < 1$. Recall that in \cite{Hogervorst:2017sfd} the one-dimensional alpha space transform was defined as
\be
\fh(\a) = \int_0^1 \frac{dz}{z^2} f(z) \Psi_\a(z)\,,
\ee
with
\be \label{1ddefinitions}
\begin{split}
\Psi_\a(z) &= \hf \left( Q(\a) k_{\a + 1/2}(z) + Q(-\a) k_{-\a + 1/2}(z)\right) = {}_2 F_1\left(\hf + \a, \hf - \a, 1, \frac{z-1}{z}\right)\,,\\
Q(\a) &= \frac{2 \G(-2\a)}{\G(\hf - \a)^2}\,,\\
k_h(z) &= z^h {}_2 F_1 (h,h,2h,z)\,.
\end{split}
\ee
The functions $\Psi_\a(z)$ have the special property that there is no (logarithmic) branch cut at $z = 1$ and have been normalized such that $\Psi_\a(1) = 1$. We claim that the $\Psi_\a(z)$ for pure imaginary $\a$ form a complete basis in a suitable function space, which implies that $f(z)$ should be recoverable from the inverse transform,
\be
f(z) = \int [d\a] \frac{2 \Psi_\a(z)}{Q(\pm \a)} \fh(\a)\,,
\ee
with the integral along the imaginary $\a$ axis. Here and below we use the conventions
\be
[d\a] = \frac{d\a}{2\pi i}\,, \qquad \qquad Q(\pm \a) = Q(\a)Q(-\a)\,,
\ee
and more generally for a function $X(\a)$ we will use $X(\pm \a)$ to denote the product $X(\a)X(-\a)$. As illustrative examples, let us list the following one-dimensional alpha space transforms:
\begin{align}
z^p \qquad & \Leftrightarrow \qquad  \frac{\G(p - \hf \pm \a)}{\G(p)^2}\\
\frac{z^p}{(1-z)^q}\qquad &\Leftrightarrow \qquad \int [ds] \frac{\G(-s) \G(s + \hf \pm \a)\G(1 -q +s)\G(p -1 -s)}{\G(s+1)\G(\hf \pm \a)\G(p-q)}\\
\left(\frac{z}{1-z}\right)^p \qquad &\Leftrightarrow \qquad   \frac{\G(1-p)\G(p-\hf \pm \a)}{\G(p) \G(\hf \pm \a)}\\
k_{\b + 1/2}(z) \qquad &\Leftrightarrow \qquad \frac{Q(-\b)}{2} \left( \frac{1}{\b - \a} + \frac{1}{\b + \a}\right)\,,
\label{ypalphatransform}
\end{align}
with the third expression obtainable as a limit of the second one. Further examples are discussed in appendix \ref{app:weird}. The last equation shows how conformal blocks are encoded as simple poles in $\fh(a)$ and this makes alpha space particularly suitable to immediately recover the conformal block decomposition of a given function $f(z,\zb)$. As an example, in appendix \ref{app:DSDformula} we use alpha space to prove the non-trivial conformal block decomposition appearing in the `remarkable exact identity' stated in equation (1.2) of \cite{Simmons-Duffin:2016wlq}.

In this note we consider functions $f(z,\zb)$ defined on the Lorentzian square (where $z$ and $\zb$ are independent and $0 < z,\zb < 1$) and obeying $f(z,\zb) = f(\zb,z)$. For these functions we define, in analogy with \cite{Hogervorst:2017sfd}, an alpha space transform as
\begin{equation}\label{D2Transform}
\boxed{
\hat{f}(\a,\bar{\a})=\int_0^1 \frac{dz}{z^2}\,\int_0^1 \frac{d\bar{z}}{\bar z^2}\, \Psi_\a(z)\Psi_{\bar{\a}}(\bar{z}) f(z,\bar{z})
}
\end{equation}
and whose inverse transform we claim to exist and be equal to
\begin{equation}\label{D2Inverse}
\boxed{
f(z,\bar{z})=\int\left[d\a\right]\int\left[d\bar{\a}\right]\frac{4 \Psi_\a(z)\Psi_{\bar{\a}}(\bar{z})}{Q(\pm \a) Q(\pm \bar{\a})}\hat{f}(\a,\bar{\a})\,.
}
\end{equation}
These are the two defining equations that provide the starting point for the analyses below.

Our functions $f(z,\zb)$ will be assumed to have a convergent two-dimensional $s$-channel conformal block decomposition given by
\be \label{D2schannelblockdec}
\begin{split}
f(z,\bar z) &= \sum_n \lambda_n^2 G_{\D_n}^{(J_n)}(z,\bar z)\,,\\
G_{\D}^{(J)} (z,\zb) &= k_{h}(z) k_{\hb}(\zb) +  k_{h}(\zb) k_{\hb}(z) \qquad \left( h = \frac{\D - J}{2}, \,\, \hb = \frac{\D + J}{2} \right)
\end{split}
\ee
and with $k_h(z)$ already defined in \eqref{1ddefinitions}. The link between this conformal block decomposition and the alpha space decomposition is the following. Since $\fh(\a,\ab)$ is even and symmetric in its arguments we may rewrite \eqref{D2Inverse} as
\be
f(z,\bar{z})=\int\left[d\a\right]\int\left[d\bar{\a}\right]\frac{k_{\a + 1/2}(z)k_{\ab + 1/2}(\bar{z})}{Q(- \a) Q(- \bar{\a})}\hat{f}(\a,\bar{\a})\,,
\ee
after which we can deform the contours of integration into the right half planes and pick up poles along the real $\a$ and $\ab$ axis. Notice that $1/Q(-\a)$ is regular for $\Re(\a) > -1/2$ so there are no kinematical poles to worry about. Instead we only get contributions from `twin poles' in the alpha space density, which are of the form
\be\label{pairofpolesinalpha}
\fh(\a,\ab) \supset \frac{R}{(\a - h + \hf)(\ab - \hb + \hf)} + (\a \leftrightarrow \ab)\,.
\ee
These give rise to a conformal block in position space,
\be
f(z,\zb) \supset \lambda^2 \left( k_h(z) k_\hb(\zb) + (z \leftrightarrow \zb) \right) = \lambda^2 G_{\hb + h}^{(\hb - h)}(z,\zb)\,,
\ee
with a coefficient
\be
\lambda^2 = \frac{4 R}{Q(-h + \hf)Q(-\hb + \hf)}\,.
\ee
We can observe the following map between $\a$ and $\ab$ and the more conventional ways of labelling a conformal representation:
\be
\begin{split}
\a &= h - \hf = \frac{\D - J -1}{2}\,,\\
\ab  &= \hb - \hf = \frac{\D + J + 1}{2}\,.
\end{split}
\ee
From its definition \eqref{D2Transform} it is clear that $\fh(\a,\ab)$ is invariant under Weyl reflections:
\be
\fh(\a,\ab) = \fh(-\a,\ab) = \fh(\a,-\ab) = \fh(\ab,\a)\,,
\ee
and therefore the two twin poles of \eqref{pairofpolesinalpha} are always part of a set of eight twin poles for every conformal block. For large enough $h$ and $\hb$ these `shadow twin poles' will sit at negative $\a$ or negative $\ab$ and therefore not be picked up if we close the integration contours in the right half planes.

Let us comment on the special status of the identity operator which sits at $\a = \ab = -1/2$. To include this operator one should add a small \emph{additional} contour to the integral where both $\a$ and $\ab$ go around $-1/2$ so they pick up the pole in $[Q(-\a)Q(-\ab)]^{-1}$ at that point, if desired with a mirror contour at $\a = \ab = 1/2$ to make the whole expression even in $\a$ and $\ab$ again. This subtlety is symptomatic of a more general problem: for physical four-point functions in unitary CFTs the integral in \eqref{D2Transform} often does not converge because of divergences along the boundary of the integration region. We refer to \cite{Hogervorst:2017sfd} for a prescription on how to deal with these issues.

\section{The lightcone bootstrap}
To gain some intuition for the alpha space transform in two spacetime dimensions we discuss how to recover the familiar lightcone bootstrap results of \cite{Komargodski:2012ek,Fitzpatrick:2012yx}.

\subsection{Large alpha behavior in one dimension}
\label{subsec:largealphaonedimension}
An important ingredient for the following discussion will be a result of \cite{Hogervorst:2017sfd} pertaining to the one-dimensional alpha transform $\fh(\a)$ of a function $f(z)$. Roughly speaking, if
\be 
f(z) = (1-z)^{-\rho}(1 + O(1-z))\,,
\ee
with $\rho$ not an integer, then
\be \label{largeimalpha}
\fh(\a) = (-\a^2)^{\r - 1}\frac{\G(1-\r)}{\G(\r)}(1 + O(\a^{-2}))\,.
\ee
for large and purely imaginary $\alpha$. Indeed, since $\Psi_\a(1) = 1$ it is simple to see that the large imaginary $\alpha$ behavior of $\fh(\a)$ is related to the $z \to 1$ limit of $f(z)$, and the precise match can then easily be recovered by considering the examples in \eqref{ypalphatransform}.

In more detail, we believe that the above statement should be understood in the same sense as the claim that the Fourier transform of a function $f(x)$ that behaves like $(x^2)^{-\D}$ for small $x$ is given by $|p/2|^{2 \D - 1} \sqrt{\pi} \G(1/2 - \D) /\G(\D) $ for large real $p$. In Fourier space subtleties arise (a) when $\D$ is a non-positive integer and we cannot conclude anything about the large $p$ expansion, (b) from delta-function-like terms supported at $x = 0$ which would correspond to the addition of arbitrary polynomials in $p$, and (c) for $\D - 1/2$ a positive integer, when the position-space distribution is singular and after suitable regularization we find logarithmic behavior in momentum space.

For the alpha space transform case (a) occurs when $\r$ is a non-positive integer and in that case we similarly cannot conlude anything about the large $\r$ behavior. Furthermore, we explain in appendix \ref{app:weird} that polynomials in alpha space correspond to delta-function-like terms supported at $z = 1$ in position space which is entirely analogous to case (b). Finally, case (c) occurs when $\rho$ is a positive integer; the logarithmic behavior of the (regularized) alpha space transform can be found by taking a limit in \eqref{ypalphatransform} and subtracting a polynomial part.

What interests us more than the large imaginary alpha behavior are the OPE coefficients, here encoded as poles which (by the assumption that $f(z)$ has a convergent conformal block decomposition with real scaling dimensions) lie along the real alpha axis. In order to investigate them it is convenient to define
\be
\text{disc}_\alpha [ \fh(\a) ] \colonequals \frac{1}{2\pi i} \lim_{\e\rightarrow 0}\left(\fh(\a + i \e) - \fh(\a - i \e)\right)\,,
\ee
which is a sum of delta functions since $\fh(\a,\ab)$ has poles on the real $\a$ axis.\footnote{For example, $\text{disc}_{\a}[\a^{-1}] =  - \delta(\a)$. In one-dimensional alpha space the residues have the opposite sign as the OPE coefficients, so this discontinuity would normally be positive.} To connect the large imaginary alpha behavior \eqref{largeimalpha} to the asymptotic behavior of $\text{disc}_\a[\fh(\a)]$ we can offer the following crude derivation. First we use a dispersion relation trick to find that\footnote{Writing a subtracted dispersion relation does not significantly change the argument. Such subtractions are necessary if $\r > 1$ or if \eqref{largeimalpha} does not hold along every non-real ray in the complex alpha plane. If $\fh(\a)$ is not polynomially bounded for large non-real $\a$ then our argument does not work; it would be interesting to show that this can never happen if $f(z)$ represents a physical correlation function.}
\be
\begin{split}
\fh(\a) &= \int_0^\infty d\b\, \frac{2\b}{\b^2 - \a^2} \text{disc}_\b [\fh(\b)]\\
		&= \int_0^\infty ds\,  e^{\a^2 s} \left( \int_0^\infty d\b\,  2 \b e^{-\b^2 s}  \text{disc}_\b [\fh(\b)] \right)\,.
\end{split}
\ee
Plugging in the behavior \eqref{largeimalpha} for large imaginary $\alpha$ we recognize that we can apply the Hardy-Littlewood Tauberian theorem (first applied to OPE data in \cite{Pappadopulo:2012jk}) to learn that
\be
\int_0^\infty d\b \, 2 \b e^{-\b s}  \text{disc}_\b [\fh(\b)] \sim \frac{s^{-\rho}}{\G(\rho)}\,,
\ee
for small $s$ and up to pieces analytic in $s$. This is a precise way of saying that, for large $\beta$,
\be
\text{disc}_\b [\fh(\b)] \approx \frac{\b^{2\rho - 2}}{\G(\rho)^2}\,,
\ee
in an averaged sense. The prefactor $1/\G(\rho)^2$ is interesting. First of all, for $\rho$ a positive integer the intermediate steps are not valid and would need regularization, but our final result is regular and we believe it accurately describes the asymptotic behavior of the OPE coefficients after all. Secondly, as before we learn that the asymptotic behavior of the discontinuity is effectively zero when $\rho$ is a non-positive integer. This time, however, the prefactor has double zeroes and therefore we can take a $\rho$-derivative of both sides to find that any $z \to 1$ behavior of the form
\be
(z-1)^k \qquad \text{or} \qquad (z-1)^k \log(z -1)\,, \qquad \text{with } k \in \{0, 1, 2, \ldots\}\,,
\ee
contributes zero to the asymptotic behavior of the discontinuity. This, of course, is precisely the behavior called `Casimir regular' introduced in \cite{Simmons-Duffin:2016wlq}. As explained there, it makes intuitive sense because this position-space behavior can easily be engineered by a finite linear combination of $s$-channel blocks, for which $\text{disc}_\b[\fh(\b)]$ would vanish identically for sufficiently large $\b$.

\subsection{Leading order}
Let us return to the two-dimensional alpha space transform. For identical operators we expect the $\zb \to 1$ limit of our correlation function to be dominated by the unit operator in the $t$ channel, which is $|z / (1-z)|^{2 \D_\phi}$. The alpha space transform of the holomorphic part is
\be \label{eq:alphatchannelid1d}
\int_0^1 \frac{dz}{z^2} \left(\frac{z}{1-z}\right)^{\D_\phi} \Psi_{\a}(z) = \frac{\G(\D_\phi - \hf \pm \a)\G(1-\D_\phi)}{\G(\hf \pm \a)\G(\D_\phi)} \,.
\ee
The large imaginary $\ab$ behavior of the alpha space density is correspondingly
\be \label{asymptotimagalpha}
\fh(\a,\ab) = (-\ab^2)^{\D_\phi - 1}\left( \frac{\G(\D_\phi - \hf \pm \a)\G(1-\D_\phi)^2}{\G(\hf \pm \a)\G(\D_\phi)^2}  + O(\ab^{-\#}) \right)\,.
\ee
In this equation we indicated how we expect the corrections to be power-law suppressed, with an exponent $\#$ that we will determine below. In terms of the discontinuity $\text{disc}_\ab[\fh(\a,\ab)]$ the behavior \eqref{asymptotimagalpha} then yields
\be
\int^\infty e^{-\ab s} \text{disc}_\ab [ \fh(\a,\ab) ] \sim \frac{\G(2\D_\phi-1)\G(1-\D_\phi)}{s^{2\D_\phi-1}\G^3(\D_\phi)} \frac{\G(\D_\phi - \hf \pm \a)}{\G(\hf \pm \a)} + s\text{-analytic,} \qquad \text{as s $\to 0$,}
\ee
which is a precise way of saying that for large $\ab$
\be
\label{discasymptotic}
\boxed{
\text{disc}_\ab[\fh(\a,\ab)] \approx \ab^{2(\D_\phi-1)} \frac{\G(\D_\phi - \hf \pm \a)\G(1-\D_\phi)}{\G(\hf \pm \a)\G^3(\D_\phi)} \,.
}
\ee
This expression shows that for asymptotically large $\ab$ the operator spectrum is supported at the double-twist values $\a = \D_\phi - \hf +n$, a result familiar from the lightcone bootstrap. Notice that this result holds only up to pieces analytic in $s$ which may come from isolated poles in $\ab$ and which correspond to the Casimir regular terms as we reviewed above.

It is perhaps worth pointing out that our derivation used only the $\zb \to 1$ behavior of the correlation function, which we claim to be related to the large $\ab$ behavior. Unlike the position-space analysis, we saw no need to take the $z \to 0$ limit as well. It is therefore natural to claim that our result remains valid in any regime where the $t$ channel identity dominates, which in particular would include the deep Euclidean regime $z, \zb \to 1$ which corresponds to $\a$ and $\ab$ both large. It would be interesting to see if alpha space can be a stepping stone for a more precise analysis of the asymptotics of the OPE data, perhaps following the ideas in \cite{Mukhametzhanov:2018zja}. Notice that our density $\fh(\a,\ab)$ does not have any kinematic poles, unlike the density defined via the Euclidean inversion formula, which might simplify the derivations.

\subsubsection{Translation to OPE Coefficients}
Equation \eqref{discasymptotic} only provides the $\ab$ discontinuity in an averaged sense. However, spins are required to be even integers in a physical correlation function of identical operators and therefore the residue of a pole in \eqref{discasymptotic} at $\a = \D_\phi - \hf + n$ for some integer $n$, as a function of $\ab$, can have poles (at most) at $\ab = \D_\phi - \hf + n + J$ for $J$ an even integer. If we assume that an operator of every allowed $J$ is present then the shape of the $\ab$-discontinuity is fixed. At very large $J$ we find, for every finite $n$, sequences of blocks with $(h,\hb) \to (\D_\phi + n, \D_\phi + n + J)$ with OPE coefficients given by
\be
\begin{split}
\lambda^2_{n,J} &\approx \left. 2 \frac{4}{Q(-\ab) Q(-\D_\phi + \hf - n)} \frac{\ab^{2(\D_\phi-1)}}{\G(\D_\phi)^2} \frac{\G(2\D_\phi + n - 1) }{n! \G(\D_\phi)^2} \right|_{\ab = \D_\phi - \hf + n + J}\\
&\approx 2^{- 2J -4 \D_\phi - 4 n + 5} J^{2 \D_\phi - 3/2} \frac{\pi  \Gamma (n+\D_\phi ) \Gamma (n+2 \D_\phi -1)}{\Gamma (\D_\phi )^4 \Gamma (n+1) \Gamma \left(n+\D_\phi -\frac{1}{2}\right)}\,.
\end{split}
\ee
To obtain this expression we computed the residues of \eqref{eq:alphatchannelid1d} and
added an extra factor of 2 because only even spins contribute. This matches, of course, the asymptotic behavior of the OPE coefficients in the mean-field solution which for $d = 2$ were obtained in \cite{Heemskerk:2009pn}. We have therefore reproduced the leading order lightcone bootstrap results of \cite{Komargodski:2012ek,Fitzpatrick:2012yx} for identical operators.

\subsection{The lightcone bootstrap at first subleading order}
Subleading terms in the large $\ab$ expansion originate from the low-twist operators in the $t$ channel. To take these into account one can introduce the `split' crossing kernel which one can think of as the alpha space density of a single t-channel block. The chiral part is defined as \cite{Hogervorst:2017sfd}\footnote{Notice that our $\D_\phi$ equals $2h$ in \cite{Hogervorst:2017sfd}. This expression is called the `split' kernel because it is the alpha space density for the single $t$-channel block $k_{\b + 1/2}(1-z)/Q(-\b)$ rather than for the symmetrized combination $\Psi_\b(1-z)$ that was considered in \cite{Hogervorst:2017sfd}.}
\begin{equation}
K_\text{split}(\a,\b|\D_\phi)\colonequals \frac{2}{Q(-\b)}\int_0^1 \frac{dz}{z^2}\left( \frac{z}{1-z}\right)^{\D_\phi}\Psi_\a(z)k_{\b + 1/2}(1-z)\,.
\end{equation}
In this paper we will mostly use a slightly different normalization, for which we introduce:
\be
\tilde K_\text{split}(\a,\b|\D_\phi)\colonequals \frac{Q(-\b)}{2} K_{\text{split}}(\a,\b|\D_\phi)\,.
\ee
In appendix \ref{app:splitkernel} we explain how the split kernel can be written in terms of Wilson functions. In two dimensions, a single $t$-channel block for a primary operator with dimension $h + \hb$ and spin $\hb - h$, which is given by
\be
f(z,\zb) \supset \lambda^2_{h,\hb} \left(\frac{z \zb}{(1-z)(1-\zb)} \right)^{\D_\phi} k_{h}(1-z) k_{\hb}(1-\zb) + (z \leftrightarrow \zb) \,,
\ee
would contribute to the $s$-channel density as
\begin{equation} \label{KsplitKsplit}
\fh(\a,\ab) \supset \lambda^2_{h,\hb} \tilde K_\text{split}(\a, h - 1/2 |\D_\phi) \tilde K_\text{split}(\bar{\a},\hb -1/2 |\D_\phi) + (h \leftrightarrow \hb)\,.
\end{equation}
We can now consider how this would yield subleading corrections to the mean field behavior. As before, we take $\ab$ large and $\a$ fixed. For large non-real values of its first argument the kernel behaves as 
\begin{equation} \label{Ksplitlargealpha}
\tilde K_\text{split}(\ab,\bb|\D_\phi) = \frac{\G(\frac{3}{2}-\D_\phi+\bb)}{\G(\D_\phi-\frac{1}{2}-\bb)}(-\ab^2)^{\D_\phi-\frac{3}{2}-\bb} \left( 1 + O(\ab^{-2}) \right)\,.
\end{equation}
This allows us to write the subleading correction to equation \eqref{asymptotimagalpha} in more detail as:\footnote{A quick recap of the variables used: $\D_\phi$ is the scaling dimension of the external operator, which we take to be a scalar. On the other hand, $h$ and $\hb$ are the quantum numbers of the $t$-channel operator of lowest twist and $\lambda^2_{h,\hb}$ the coefficient of the corresponding conformal block.}
\be
\begin{split}
\fh(\a,\ab) &= (-\ab^2)^{\D_\phi - 1} \frac{\G(\D_\phi - \hf \pm \a)\G(1-\D_\phi)^2}{\G(\hf \pm \a)\G(\D_\phi)^2} \\ &  + (-\ab^2)^{\D_\phi - h - 1}  \lambda^2_{h,\hb} (1 + \d_{h,\hb})  \frac{\G(1-\D_\phi+ h)}{\G(\D_\phi- h)}  \tilde K_\text{split}(\a,\hb -1/2|\D_\phi)\\ &+ O(\a^{-\#}).
\end{split}
\ee 
We in particular see that the relative power in $\alpha$ between the leading and subleading term equals the twist $2h$ of the first t-channel operator, in agreement with well-known lightcone bootstrap results. The Kronecker delta arises because both terms in \eqref{KsplitKsplit} contribute equally for a spin $0$ operator. As explained above, the asymptotic behavior at large imaginary $\a$ corresponds to a discontinuity at large real $\a$ obtained by making the replacement
\be
(-\ab^2)^{x-1}\rightarrow \frac{\ab^{2x - 2}}{\G(x)\G(1-x)}\,,
\ee
giving the correction to \eqref{discasymptotic}
\be
\label{discasymptoticsubl}
\begin{split}
\text{disc}_\ab [\fh(\a,\ab)] &\approx \ab^{2\D_\phi-2} \frac{\G(\D_\phi - \hf \pm \a)\G(1-\D_\phi)}{\G(\hf \pm \a)\G^3(\D_\phi)} \\ &  + \ab^{2\D_\phi - 2h - 2}   \frac{\lambda^2 (1 + \d_{h,\hb})}{\G(\D_\phi- h)^2}  \tilde K_\text{split}(\a,\hb - 1/2|\D_\phi)
\\ & + O(\a^{-\#})\,.
\end{split}
\ee 

To work out what this means for the OPE data we need the following details about $K_\text{split}(\a,\b|\D_\phi)$. For fixed $\b$ it has double poles at $\a = \pm ( \D_\phi - \hf + n)$, so it takes the form
\be
\tilde K_\text{split}(\a,\b|\D_\phi) = \frac{M^{(K)}_n(\b|\D_\phi)}{(\a - \D_\phi + \hf - n)^2} + \frac{N^{(K)}_n(\b|\D_\phi)}{\a - \D_\phi + \hf - n} + \text{regular}\,,
\ee 
with coefficients $M^{(K)}_n(\b | \D_\phi)$  and $N^{(K)}_n(\b | \D_\phi)$ for which we provide a closed-form expression in terms of Wilson functions in equation \eqref{mninsplitkernel}.

To fix ideas in the following it helps to think of $f(\a,\ab)$ as having the schematic form, for $\a$ near the $n$'th of the double twist poles,
\be \label{postulatedfh}
\fh(\a,\ab) \approx \sum_J \frac{R_n(J)}{(\a - \D_\phi + \hf - n - \hf \g_n(J))(\ab - \D_\phi  + \hf - n - \hf \g_n(J) - J)}\,,
\ee
which means that for each $n$ we have a sequence of conformal blocks of increasing $J$ with the following scaling dimensions and spin:
\be
\begin{split}
\Delta_n(J) &= J + 2 \D_\phi + 2n + \gamma_n(J)\,,\\
\lambda_n(J) &= \frac{4 R_n(J)}{Q(- \D_\phi + \hf - n - \hf \g_n(J) - J)Q(- \D_\phi  + \hf - n - \hf \g_n(J))}\,.
\end{split}
\ee
For the discontinuity this would imply that, in a smoothed out sense,
\be \label{eq:assumeddiscform}
\begin{split}
\text{disc}_\ab [\fh(\a,\ab)]|_{\ab = \D_\phi - 1/2 + n + \hf \g_n(J) + J} \approx - \frac{R_n(J)}{\a - \D_\phi + \hf - n - \hf \g_n(J)}\,.
\end{split}
\ee
Comparing this to \eqref{discasymptoticsubl}, and taking into account the proper change of variables as discussed in appendix \ref{app:alphatoJ}, we read off that
\be \label{Rgammalo}
\begin{split}
R_n(J) [\ab'(J)]^{-1} &\sim \ab(J)^{2\D_\phi - 2} \left( \frac{\G(2\D_\phi +n -1)}{\G(\D_\phi)^4 n!} - \ab(J)^{- 2 h} \frac{\lambda^2 (1 + \d_{h,\hb})}{\G(\D_\phi-h)^2} N^{(K)}_n(\hb - 1/2|\D_\phi)  + \ldots \right)\\
\g_n(J) &\sim  - \ab(J)^{-2h} \lambda^2 \Gamma_n(h,\hb | \D_\phi ) + \ldots
\end{split}
\ee
with $\ab(J)$ defined implicitly as the solution to
\be
\ab(J) = J + \D_\phi - \hf + n + \hf \g_n(\ab(J))
\ee
and with
\be
\Gamma_n(h,\hb | \D_\phi) =  \frac{(1 + \d_{h,\hb})}{\G(\D_\phi- h)^2} M_n^{(K)}(\hb - 1/2|\D_\phi) \frac{ \G(\D_\phi)^4 n!}{\G(2\D_\phi + n - 1)}\,.  
\ee
yielding the leading order corrections (for any $n$) to the double-twist alpha space density at large spin originating from a single $t$-channel block. For $n = 0$ we find for example that
\be
\Gamma_0(h,\hb | \D_\phi) =  (1 + \d_{h,\hb}) \frac{\G(\D_\phi)^2}{\G(\D_\phi- h)^2}\frac{\Gamma(2\hb)}{\G(\hb)^2} \,,
\ee
which matches the results in \cite{Komargodski:2012ek,Fitzpatrick:2012yx}.

\subsubsection{Reciprocity}
The subleading terms in \eqref{discasymptoticsubl} arise both from further $t$-channel blocks, on which we will comment below, and also because of subleading terms in the large $\alpha$ expansion of \eqref{KsplitKsplit}. The split kernel has an expansion in integer powers of $-\alpha^2$ relative to the leading term in \eqref{Ksplitlargealpha}. Going through the motions again, we find that the $t$-channel identity combined with \emph{just a single $t$-channel block} leads to residues and anomalous dimensions with a large $\ab$-expansion of the schematic form
\[
\begin{split}
R_n(J(\ab)) J'(\ab) &\sim \# \ab^{2\D - 2} + \ab^{2\D - 2 - 2h} \left( \# + \# \ab^{-2} + \#  \ab^{-4} + \ldots \right) \,,\\
\g_n(J(\ab)) &\sim \# \ab^{-2h} \left( \# + \# \ab^{-2} + \#\ab^{-4} + \ldots \right)\,.
\end{split}	
\]
The expansion in integer powers of $\ab^{-2}$ corresponds to the `reciprocity principle' for the double-twist operators that was highlighted in \cite{Alday:2015eya}. The expansion in even powers of $\alpha$ then becomes an expansion in even powers of the square of the right-hand side. For $n = 0$ this is (up to a constant shift) precisely the two-dimensional version of the `Casimir' defined in \cite{Alday:2015eya}, whereas the case $n > 0$ was not discussed in \cite{Alday:2015eya}.

\subsubsection{Constraints from exponentiation}
The double poles in $\a$ arising from the split kernel lead to `anomalous dimensions' in the large spin expansions. We formalized this above by postulating a form of the discontinuity in \eqref{eq:assumeddiscform}. This form gives the requisite poles upon expansion for small $\gamma$, allowing us to determine these anomalous dimensions as written in \eqref{Rgammalo}. However if \eqref{eq:assumeddiscform} is correct then the higher-order terms in the small $\gamma$ expansion require higher-order poles in $\a$. More precisely, one would expect terms like
\be
\text{disc}_\ab[f(\a,\ab)] \supset - \ab^{2 \D_\phi -2} \frac{\G(2\D_\phi +n -1)}{\G(\D_\phi)^4 n!} \sum_{k = 0}^\infty \frac{(-1)^k \ab^{- 2 k\bb - k} \lambda^k \Gamma_n^k(\b,\bb| \D_\phi)}{2^k ( \a - \D_\phi + \hf -n)^{k+1}}\,,
\ee
with poles of order $k+1$ in $\ab$ at the double-twist values. A single $t$-channel block only gives double poles, so this exponentiation of the anomalous dimension is not automatic. What would $\fh^t(\b,\bb)$ have to look like to reproduce these singularities?

As was also already explained in \cite{Simmons-Duffin:2016wlq}, the power of $\ab$ in the above expansion indicates that such operators come from t-channel operators with twists that equals $k h$. In other words, the exponentiation of the s-channel anomalous dimensions due to a single t-channel operator $\op_t$ of twist $h$ requires the existence of further t-channel operators with twist $k h$, so the multi-twist versions of the original t-channel operator. A single such multi-twist t-channel operator will not produce the higher-order pole in $\ab$ that the above expression requires, and therefore we actually need an infinite family of such multi-twist t-channel operators. If we think of $\op_t$ as ``single-particle exchange'', then a $k$-particle exchange diagram is necessary to reproduce the $(k+1)$-th order pole in the above expression. Notice furthermore that this $k$-particle exchange must somehow conspire to give the corresponding higher-order pole for all $n$.

This discussion can be quantified. One way to do so is to use the `quadruple discontinuity' as in \cite{Caron-Huot:2017vep} but we can also do it easily in alpha space: we just need to apply the crossing kernel to the exponentiated $\alpha$ space $s$-channel density, like the one postulated in \eqref{postulatedfh}, in order to find out the required $t$-channel density. Up to unimportant subtleties related to the small $\alpha$ behavior, we have in the conventions of \cite{Hogervorst:2017sfd} that
\[
\int_{- i \infty}^{i \infty} \frac{[d\a]}{N(\a)} K(\b,\a|\D_\phi) \fh(\a) = \int^\infty_0 d\a \, K_{\text{split}}(\b,\a|\D_\phi) \text{disc}_\a[ \fh(\a) ]\,,
\]
and therefore we can obtain the requisite $t$-channel density as
\[
\begin{split}
f^t(\b,\bb) \supset &\phantom{=}\int^\infty d\ab\, K_{\text{split}}(\bb,\ab|\D_\phi) \int [d\a] K_{\text{split}}(\b,\a|\D_\phi)  \left( - \frac{R_n(\ab)}{\a - \D_\phi + \hf - n - \hf \g_n(\ab)} \right) 
\\
&= \int^\infty d\ab \, K_{\text{split}}(\bb,\ab|\D_\phi) R_n(\ab) K_{\text{split}}(\b,\D_\phi - 1/2 + n + \g_n(\ab)/2|\D_\phi)\,,
\end{split}
\]
plus another term that swaps $\b$ and $\bb$. We expect the form of the integrand to only be valid for large $\ab$, so we will also substitute the large $\ab$ behavior of $K_{\text{split}}(\bb,\ab|\D_\phi)$. For large values of its second argument the split kernel behaves as
\be
K_\text{split}(\bb,\ab|\D_\phi) \sim  Q(-\bb)\G^2(\D_\phi - 1/2 -\bb)(\ab^2)^{1 + \bb - \D_\phi } + (\bb \leftrightarrow - \bb)\,.
\ee
We will also take $\b$ large in order to find an estimate at large spin in the $t$-channel. Using \eqref{Ksplitlargealpha} we then find leading order behavior of the form
\be
\begin{split}
&f^t(\b,\bb) \supset   Q(-\bb)\G^2(\D_\phi - 1/2 -\bb) \times \\ & \quad \int^\infty d\ab \,  R_n(\ab)\frac{2\G(1 + n + \g_n(\ab)/2)}{\G( -  n - \g_n(\ab)/2) Q(1/2 - \D_\phi - n - \g_n(\ab)/2)} (\ab^2)^{1 + \bb - \D_\phi } (-\b^2)^{- 1 -  n - \g_n(\ab)/2}\,.
\end{split}
\ee
As a consistency check, notice that if $\g_n(\ab)$ were to vanish then there is nothing to exponentiate, and in that case the above expression also does not teach us anything about the $t$-channel OPE density. We see that the leading trajectory with $n=0$ dominates, so let us focus on that one from now on. Passing to the discontinuity, we obtain
\be
\text{disc}_\b [f^t(\b,\bb)] \supset Q(-\bb)\G^2(\D_\phi - 1/2 - \bb) \int^\infty d\ab  \, \frac{R_0(\ab) (\ab^2)^{1 + \bb - \D_\phi} \b^{-2 - \g_0(\ab)}  }{\G( - \g_0(\ab)/2)^2 Q(1/2 - \D_\phi - n - \g_n(\ab)/2)}\,.
\ee
This is a useful intermediate expression in itself, for example if one has estimates for $\g_0(\ab)$ and $R_0(\ab)$ beyond the leading order. It is similar to equation (4.23) in \cite{Caron-Huot:2017vep}. For illustrative purposes let us substitute the leading-order behaviors of these functions and expand at large $\ab$. We find, in the leading log approximation, the multi-twist poles with residues as in
\be
\frac{\text{disc}_\b [f^t(\b,\bb)]}{Q(-\bb)\G^2(\D_\phi - 1/2 - \bb) } \supset - \frac{1}{Q(1/2 -\D_\phi)} \sum_{n = 0}^\infty \frac{(\lambda^2 \Gamma_0(h,\hb|\D_\phi))^{n + 2}}{\bb + 1/2 - h (n + 2)} \frac{\log^n(\b)}{8\b^2}\,.
\ee
In addition we also have corrections to the double twist operators from the double poles of the gamma function at $\bb = \D_\phi - 1/2 + n$. It would be useful to compute these in order to find the correction of a Regge trajectory to itself. However for this the large $\ab$ expansion that we used does not suffice. This is because for large $\b$ the dominant contribution comes from the smallest possible $\g_0(\ab)$, and since $\g_0(\ab)$ is generally negative the most important contribution will come from small $\ab$ rather than from the large $\ab$ region. On the other hand, the new multi-twist poles at $\bb = h(n+2) -1/2$ that we have just described can only appear from the large $\ab$ region.

\subsection{On the convergence of the lightcone bootstrap}
Incorporating further $t$-channel operators leads one to consider an equation of the form
\be \label{KsplitKsplitsum}
\fh(\a,\ab) \overset{?}{=} \sum_{k} R_k \left( K_\text{split}(\a,\b_k | \D_\phi) K_\text{split}(\ab,\bb_k|\D_\phi) + (\b_k \leftrightarrow \bb_k) \right)\,,
\ee
with $k$ labelling the different $t$-channel primary operators. Unfortunately the convergence of the above expansion is problematic. For example, suppose we would like to reproduce the $s$-channel identity. As we have just seen (albeit with $s$- and $t$-channel swapped), this fixes the asymptotic form of the discontinuity as in \eqref{discasymptotic} which effectively says that we need an infinite set of $t$-channel blocks with ever-increasing spin $J = \bb - \b$, with twists $2 \b + 1$ approaching $ 2 \D_\phi$, and with residues $R_J$ scaling like $J^{2\D_\phi - 3/2}$. However, as we already stated above, for large $\bb$ the kernel behaves as
\be \label{largebetakernel}
K_\text{split}(\ab,\bb|\D_\phi) \sim \left( Q(-\ab)\G^2(\D_\phi - \hf -\ab)(\bb^2)^{1 + \ab - \D_\phi } + (\ab \leftrightarrow - \ab) \right)\,,
\ee
where for $\Re(\ab) \neq 0$ the leading behavior is given by picking the dominant power on the right-hand side. Substituting $\bb \sim J$ and $R_J \sim J^{2 \D_\phi - 3/2}$ we find $J^{\pm 2 \ab}$ as the net behavior, and have to conclude that the sum over $J$ converges \emph{nowhere} in the complex $\ab$ plane. In short, it suffices to consider just a single Regge trajectory to show that the sum in \eqref{KsplitKsplitsum} cannot converge anywhere.\footnote{A finite window of convergence could be obtained if the $s$-channel identity operator were absent, for example if the operators were not pairwise identical. One can also by hand subtract the $s$-channel identity, written as an infinite sum of $t$-channel blocks, which would result in a conditionally convergent sum. It would be interesting to work this out further.}

We can try to remedy this non-convergence as follows. Let us introduce a ``doubly split'' kernel, defined as
\be \label{dsplitkernel}
K_\text{dsplit}(\ab,\bb|\D_\phi) = \frac{Q(\ab)}{Q(-\bb)} \int_0^1 \frac{dz}{z^2} \left(\frac{z}{1-z}\right)^{\D_\phi} k_{\ab + 1/2}(z) k_{\bb + 1/2}(1-z)\,.
\ee 
For positive real $\bb$, this split kernel is analytic in the right half of the $\ab$ plane except for the kinematical simple poles of $Q(\ab)$ at positive integer $\ab$. The sum \eqref{KsplitKsplitsum} can then be rewritten as\footnote{Recall that for physical operators $\bb_k = \hf (\D_k + J_k - 1) \geq \hf(\D_k - J_k - 1) = \b_k$ so $\b_k$ can only grow large if $\bb_k$ does as well.}
\be \label{KdsplitKsplitsum}
\fh(\a,\ab) \overset{?}{=} \sum_{k} R_k K_\text{split}(\a,\b_k | \D_\phi) K_\text{dsplit}(\ab,\bb_k|\D_\phi) + (\ab \leftrightarrow - \ab) + (\a \leftrightarrow \ab)\,.
\ee
The large $\bb$ limit of the doubly split kernel retains only one of the two powers in \eqref{largebetakernel}:
\be \label{largebetadsplitkernel}
K_\text{dsplit}(\ab,\bb|\D_\phi) \sim Q(\ab)\G^2(\D_\phi - \hf  + \ab)(\bb^2)^{1 - \ab - \D_\phi }\,.
\ee
Returning to the analysis of a single Regge trajectory, we now find that the summand in \eqref{KdsplitKsplitsum} behaves as $J^{-\ab}$ for large $J$ and therefore converges as long as $\Re(\ab) > 1$. Our earlier objection against the convergence of the full sum in \eqref{KdsplitKsplitsum} is therefore neutralized.

It is worthwhile to pursue the analysis of the convergence in \eqref{KdsplitKsplitsum} a bit further. The $s$-channel identity operator has a $t$-channel density takes the form
\be
\frac{\G(\D_\phi - \hf \pm \b)\G(1-\D_\phi)}{\G(\hf \pm \b)\G(\D_\phi)} \times (\b \leftrightarrow \bb)\,,
\ee
and picking up the right poles equation \eqref{KdsplitKsplitsum} then results in the following putative expression for the alpha space transform $\fh_{\textbf 1} (\a,\ab)$ of the $s$-channel identity operator:
\be \label{eq:fh1putative}
\begin{split}
\fh_{\textbf 1} (\a,\ab) \overset{?}{=} 
&\sum_{m = 0}^\infty \sum_{n = 0}^m \frac{\G(2\D + n -1) \G(2\D + m -1)}{\Gamma(\D)^4(1 + \d_{m,n})m! n!}
\\&\qquad \times K_\text{split}(\a, \D_\phi + n -1/2 | \D_\phi) K_\text{dsplit}(\ab,\D_\phi + m - 1/2|\D_\phi)
\\ &+ (\ab \leftrightarrow - \ab) + (\a \leftrightarrow \ab)\,.
\end{split}
\ee
For large $m$ and $n$ the summand behaves as $n^{2 \pm \a} m^{- 2\ab}$ and therefore the sum converges if
\be \label{analyticityregion}
\Re(\ab) > \pm \Re(\a) + 1\,.
\ee
It is now natural to conjecture that a more realistic sum over $t$-channel blocks will still converge in this wedge. If this is the case then we recover there a function with purely kinematical poles: simple poles in $\ab$ from the prefactor $Q(-\ab)$ in the doubly split kernel \eqref{dsplitkernel} and double poles in $\a$ at the double-twist values from the split kernel itself. If we call this function $\fh_{\text{an}}(\a,\ab)$ (with `an' for analytic) then the full $\fh(\a,\ab)$ is recovered from this function as
\be
\fh(\a,\ab) = \fh_\text{an}(\a,\ab) + \fh_\text{an}(\a,-\ab) + (\a \leftrightarrow \ab)\,.
\ee
Clearly this requires going outside the region \eqref{analyticityregion} and this is how we envisage that the kinematical poles in $\fh_\text{an}(\a,\ab)$ get replaced by physical poles in $\fh(\a,\ab)$.

In appendix \ref{app:almostidentity} we continue our analysis of $\fh_{\textbf 1} (\a,\ab)$ and explain how the sums in equation \eqref{eq:fh1putative} allow us to almost (!) recover the constant function from a sum over $t$-channel blocks. In section \ref{subsec:tchannelsumanalytic} we offer some more comments on using truncated sums in equations \eqref{KsplitKsplitsum} and \eqref{KdsplitKsplitsum} as approximations for the $s$-channel density.

\section{A more analytic density}
\label{sec:moreanalytic}
The alpha space density $\fh(\a,\ab)$ is inspired by, but not the same as, another density that was introduced about forty years ago in \cite{Dobrev:1977qv} and obtained from harmonic analysis on the conformal group. Let us call this latter density a `Euclidean' density and denote it as $c(\a,\ab)$. This density has received much attention recently because it can be computed using the so-called `Lorentzian inversion formula' of \cite{Caron-Huot:2017vep} which manifests analyticity in spin (in some domain).

The most important difference between the Euclidean density and the alpha space density is in the way conformal blocks are encoded by singularities. In the Euclidean density $c(\a,\ab)$, a conformal block $G_{\D}^{(J)}(z,\zb)$ (with unit coefficient) originates from a pole like
\be
c(\a,\ab) \supset \left. \frac{ (1 + \d_{J,0})}{\k(\ab)} \times \frac{1}{\a + \ab - \D}  \right|_{\ab = \a + J}\,,
\ee
whereas in alpha space it would be encoded by twin poles as
\be
\fh(\a,\ab) \supset \frac{\frac{1}{4} Q(-\a)Q(-\ab)}{(\a - (\D - J - 1)/2)(\ab - (\D + J -1)/2)} + (\a \leftrightarrow \ab)\,.
\ee
This fundamental difference makes it not straightforward to link the two densities. However, using our knowledge of the alpha space density it is easy to introduce a slightly different density which also only has single poles, much like $c(\a,\ab)$. Starting from this density, then, we can fairly easily obtain the Lorentizan inversion formula (in two dimensions). This (somewhat heuristic) derivation of the Lorentzian inversion formula does not proceed in the usual sense, which is by demonstrating its equivalence to the Euclidean inversion formula \cite{Caron-Huot:2017vep,Simmons-Duffin:2017nub,Kravchuk:2018htv} (see also \cite{Isachenkov:2017qgn}), but rather by showing that whatever density it produces indeed has poles that encode the physical spectrum. To avoid clutter we will first focus only on the $z$ dependence -- we will reinstate the $\zb$ dependence in subsection \ref{subsec:twovars}.

Consider, then, a single-variable function $f(z)$. We will suppose that it has an \emph{integer-spaced} conformal block decomposition of the form
\be \label{intspacing}
f(z) = \sum_{m = 0}^{\infty} \lambda_{m} k_{p + m} (z)
\ee
which converges for $z$ between $-\infty$ and $1$. In the alpha space transform of $f(z)$,
\be
\fh(\a) = \int_0^1 \frac{dz}{z^2}\Psi_\a(z) f(z)\,,
\ee
we find poles when $\a = p + m - 1/2$ with residue $R_m = - \lambda_m Q(-p - m + 1/2)/2$. These poles arise because for these values of $\a$ the small $z$ expansion of the integrand contains a $z^{-1}$ term, and the residue $R_m$ of the alpha space pole is just the coefficient of this term. If there are no other terms that cause branch cuts at the origin, which is the case if we assume an integer-spaced conformal block decomposition and if we remove the $k_{\a + 1/2}(z)$ bit from $\Psi_{\a}(z)$, then we can get $R_m$ also by simply contour integrating around the origin. Indeed, one may check that, for any $h$,
\be \label{resblockblock}
\text{Res}_{z = 0}\left[ \frac{1}{z^2} k_{h + m}(z)k_{1 - h - n}(z) \right] = \delta_{n,m}\,,
\ee
which is a formula that was already used in \cite{Heemskerk:2009pn}.

By putting in all the right factors and deforming the integration contour somewhat this means that for the density $\ft(\a)$ defined as
\be \label{ft}
\ft(a) \colonequals - \int_{-\infty}^{0^+} \frac{dz}{z^2} \text{disc}_{z}[f(z) k_{-\a + 1/2}(z) ]
\ee
we can recover the OPE coefficients on the nose by simply evaluating it at the right $\ab$:
\be \label{lambdafromft}
\lambda_m = \ft(p +m -1/2) \,.
\ee
In contrast with $\fh(\a)$, there is no need to take a residue.

Notice that the discontinuity will (by construction) have delta function contributions as $z = 0$ which of course need to be taken into account - this is why we wrote $0^+$ as the upper limit of the integral. A more accurate definition would be that
\be
\ft(\a) \colonequals \frac{1}{2\pi i} \int_C \frac{dz}{z^2}  f(z) k_{- \a + 1/2}(z)\,,
\ee
where $C$ is a Hankel contour that starts at $-\infty - i \e$, runs parallel to the real axis until it wraps around the origin in a counterclockwise fashion, and then extends to $- \infty + i\e$.

From its definition we observe that $\ft(\a)$ is well-defined as long as $f(z) < O(z^{1-\e})$ as $z \to - \infty$ (both slightly above and below the real axis). We will assume this to be the case for now, although in the physical case (and for large external dimensions) some subtractions may be necessary.

The previous discussion needs refinement when $p - 1/2 \in \mathbb Z$. This is because the `shadow' conformal blocks have poles for positive integer values of $\a$, that is for $m \in \mathbb N^+$ we have
\be
k_{-\a + 1/2}(z) = \frac{r_m k_{m + 1/2}(z)}{\a - m}  + O(1)\,, \qquad  r_m \colonequals \frac{2^{-4 m}\pi}{\Gamma \left(\frac{1}{2}-m\right)^2 \Gamma (m) \Gamma (m+1)}\,.
\ee 
The definition \eqref{ft} therefore becomes singular and indeed generically does have poles for negative integer $\a$. For $p$ a positive half-integer, however, equation \eqref{lambdafromft} instructs us precisely to evaluate $\ft(\a)$ at these singular values. As it turns out this singularity is removable precisely in these cases so one can define $\ft(\a)$ at integer $\a$ via the limit. This does not completely resolve the issue: for half-integer $p$ the OPE coefficients are now given by:
\be \label{lambdafromftspecial}
\lambda_m = \ft(p + m - 1/2) - r_{p + m - 1/2} \ft'(- p - m + 1/2) \qquad \text{(for $p-1/2 \in \mathbb Z$).}
\ee
as opposed to direct evaluation as in \eqref{lambdafromft}. We emphasize that the first term is defined through the limit, and the second term involves a derivative.

\begin{figure}
\begin{center}
\includegraphics[width=9cm]{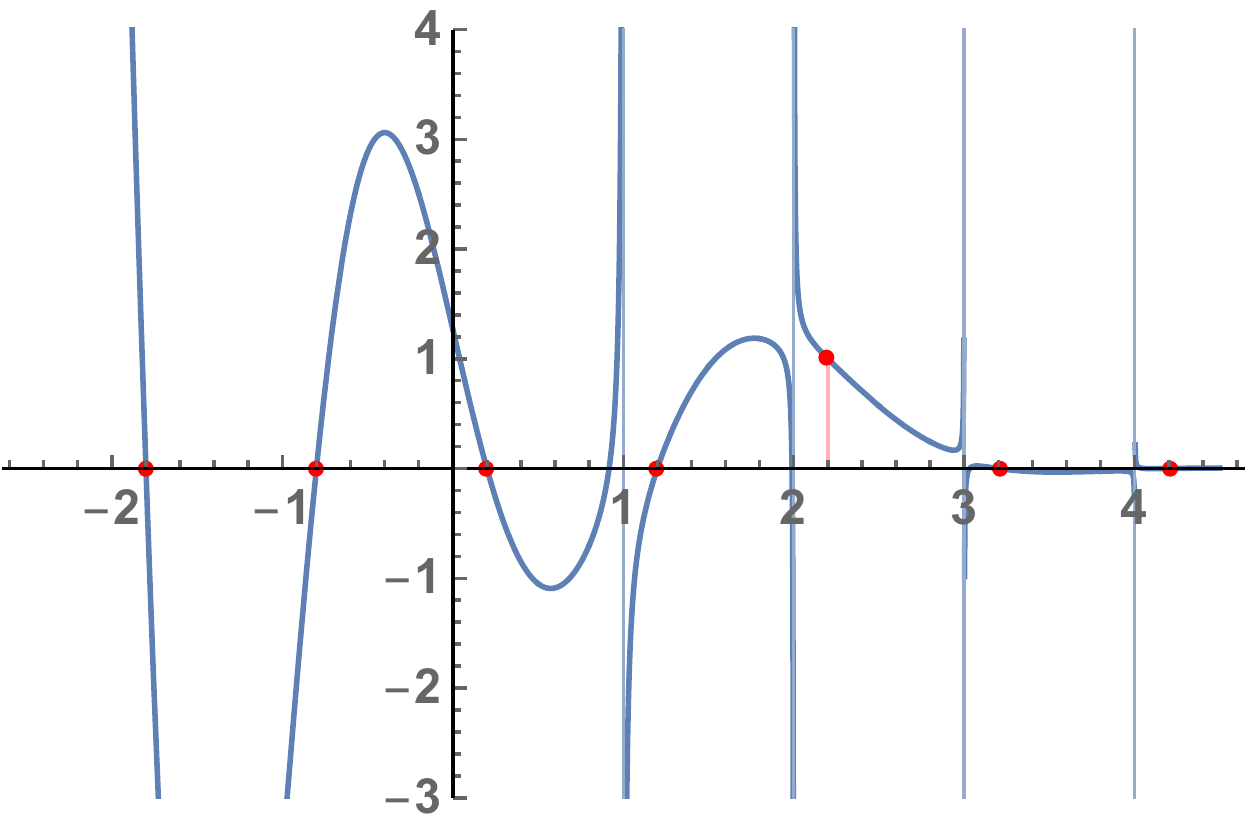}
\end{center}
\caption{\label{fig:JCab}A plot of $J^{(C)}(\a,\b)$ with $\b = 2.2$ as a function of (real) $\alpha$. We observe the kinematic poles at positive integer $\alpha$. The red dots help to show that $J^{(C)}(\b + n,\b) = \delta_{n,0}$.}
\end{figure}

It is instructive to consider the density for a single $s$-channel block. We define
\be
J^{(C)}(\a,\b) \colonequals \int_C \frac{dz}{z^2} k_{-\a + 1/2}(z) k_{\b + 1/2}(z)\,,
\ee
which is the `analytic' continuation of the Kronecker delta in \eqref{resblockblock} in the sense that $J^{(C)}(\b + n,\b) = \delta_{n,0}$. It is analytic in both $\alpha$ and $\beta$ except for the  poles at positive integer $\alpha$ and negative integer $\beta$. For $\beta = 2.2$ we plot $J^{(C)}(\a,\b)$ in figure \ref{fig:JCab}.

\begin{figure}
\begin{center}
\includegraphics[width=9cm]{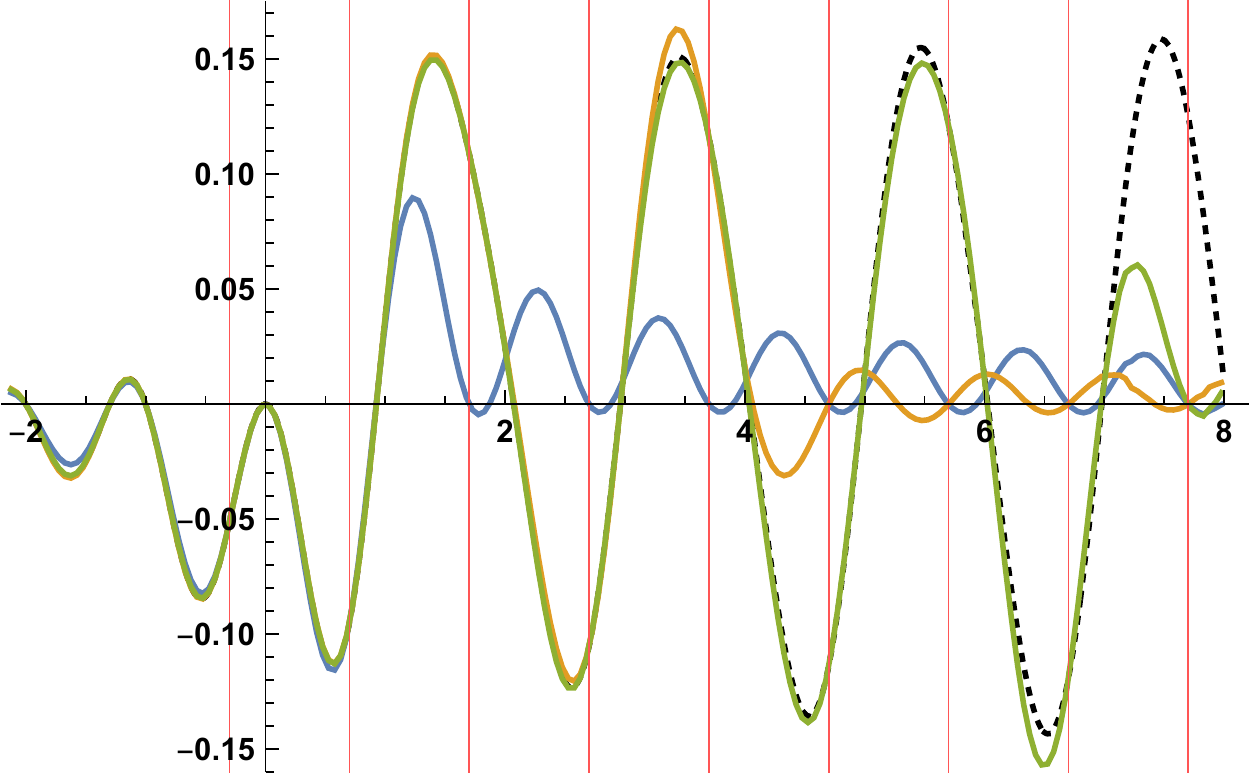}
\caption{\label{fig:schanneldec}Convergence of the $s$-channel block decomposition for the transform of $z^p(1-z)^{-q}$ with $p = 0.2$ and $q = 0.3$. The dotted line corresponds to $I^{(C)}_{p,q}(\a)$. The blue, orange and green lines are obtained by keeping respectively 2, 5 and 8 terms in the sum on the right-hand side of equation \eqref{ICpqschannelblocks}. Notice that they respectively agree with the exact result for 2, 5 and 8 non-negative integer values of $\a - p + 1/2$ which are indicated by the red vertical lines. (For clarity of presentation we have divided all functions by $4^{-\a} \G(\pm \a)$ which in particular removed all the poles at positive integer $\alpha$.)}
\end{center}
\end{figure}

If the $z \to \infty$ behavior is sufficiently benign then the $s$-channel block decomposition commutes with doing the integral along $C$. For example, if we define
\be
I^{(C)}_{p,q}(\a) \colonequals \int_C \frac{dz}{z^2} k_{-\a + 1/2}(z) z^p (1-z)^{-q}\,,
\ee
which converges for $p-q < 1$, then we would expect that
\be \label{ICpqschannelblocks}
I^{(C)}_{p,q}(\a) = \sum_{m = 0}^\infty I^{(C)}_{p,q}(p + m - 1/2) J^{(C)}(\a, p + m - 1/2)\,,
\ee
with the right-hand side a convergent sum. In figure \ref{fig:schanneldec}, we show how truncated sums on the right-hand side approximate the left-hand side in a specific example.

\begin{figure}
\begin{center}
\includegraphics[width=9cm]{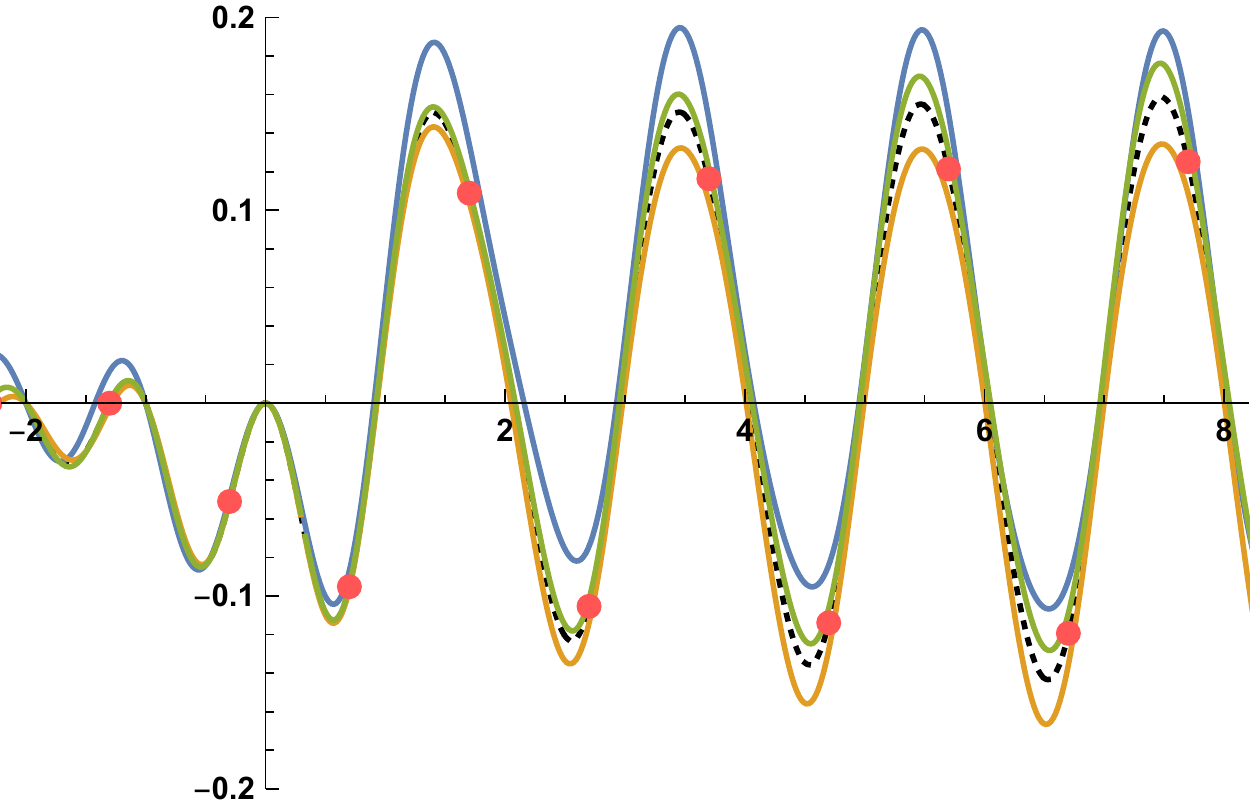}
\caption{\label{fig:tchanneldec}Convergence of the $t$-channel block decomposition for the transform of $z^p(1-z)^{-q}$ with $p = 0.2$ and $q = 0.3$ and with $h = 0.7$. The dotted line corresponds to $I^{(C)}_{p,q}(\a)$, and the red dots highlight the values at $\alpha = p - 1/2 + m$, $m$ integer, which are the OPE coefficients. The blue, orange and green lines are obtained by keeping respectively 2, 5 and 8 terms in the sum on the right-hand side of equation \eqref{ICpqtchannelblocks}. (For clarity of presentation we have divided all functions by $4^{-\a} \G(\pm \a)$ which in particular removed all the poles at positive integer $\alpha$.)}
\end{center}
\end{figure}

We can also consider the density for a single $t$-channel block, which is a crossing kernel of sorts. We set
\be
K^{(C)}_h(\a,\b) \colonequals \int_C \frac{dz}{z^2} k_{-\a + 1/2}(z) k_{\b + 1/2}(1-z) \left(\frac{z}{1-z}\right)^h\,,
\ee
which differs from the doubly split kernel introduced above by its choice of integration contour. In isolation this object is slightly  difficult to interpret because a $t$-channel block cannot be decomposed into $s$-channel blocks, but we should be able to recover the $s$-channel OPE data from an infinite sum of such transformed $t$-channel blocks. For the function $z^p (1-z)^{-q}$ this for example leads us to the statement that
\be \label{ICpqtchannelblocks}
I^{(C)}_{p,q}(\a) = \sum_{m = 0}^\infty I^{(C)}_{h-q,h-p}(h - q + m - 1/2) K^{(C)}_h(\a, h - q + m - 1/2)\,.
\ee
Here we used that its $t$-channel decomposition consists of conformal blocks with dimensions $h - q + m$ and with OPE coefficients given by $I^{(C)}_{h-q,h-p}(h - q + m - 1/2)$. In figure \ref{fig:tchanneldec}, we again show how truncated sums on the right-hand side approximate the left-hand side in a specific example. Remarkably we observe that convergence works best for $\a$ around zero -- it would be nice to have a rigorous derivation of this result.

It is of interest to consider the convergence of the $t$-channel block decomposition, for which we would need to know the large $\b$ behavior of $K^{(C)}_h(\a,\b)$. Although we were unable to obtain an exact result, some numerical experimentation leads us to believe that
\be
K^{(C)}_h(\a,\b) \sim \frac{Q(-\b)}{\b}\qquad\text{as }\b \to \infty\,,
\ee
with an undetermined prefactor that depends on $\alpha$ and $h$. On the other hand, we know from the analysis in section \ref{subsec:largealphaonedimension} that the OPE coefficient density behaves as $\b^{2 h - 2} / Q(-\b)$ so as to reproduce the identity operator in the other channel. We therefore find the decomposition of $\ft(\a)$ into a sum over $t$-channel blocks should be convergent as long as $h < 1$. This was to be expected, since the contour $C$ lies in the region where both the $s$ channel and the $t$ channel block decomposition converge and for $h < 1$ there are no issues at infinity.

\subsection{Relation to double discontinuities}
By pulling the rightmost endpoint of the contour a bit more rightward we may write 
\be \label{ftcontourto1}
\ft(\a)  = \frac{1}{2\pi i} \int_{-\infty}^1 \frac{dz}{z^2} \left( f(z - i \e) k_{- \a + 1/2}(z - i \e) - f(z + i \e) k_{- \a + 1/2}(z + i \e)\right)\,,
\ee
where we recall that we are considering functions $f(z)$ with integer-spaced conformal block decomposition,
\be
f(z) = \sum_{m = 0}^\infty \lambda_m k_{p + m}(z)\,,
\ee
for some real $p$. The astute reader may have observed the missing $i \e$ prescription for the measure $1/z^2$ in \eqref{ftcontourto1}. In fact, the correct prescription is always the same as indicated in the remainder of the integrand and we will simply continue to omit it to avoid clutter.

For the following discussion to go through we will need to assume a few things. First of all we will suppose that the singularities in $f(z)$ at $z = 1$ and $z = \infty$ are not too severe, in the sense that $f(z) = O((1 -z)^{-1 + \e})$ as $z \to 1$ and $f(z) = O(z^{-\e})$ as $z \to \infty$ for some $\e > 0$. Furthermore, $f(z)$ has a `kinematic' cut from $0$ to $-\infty$; let us suppose that $f(z)$ further only has a cut from $1$ to $+ \infty$ and that the structure of cuts on the secondary sheets that we find by passing through either of these two cuts is the same. These assumptions appear to be satisfied if $f(z)$ is a physical correlation function in a one-dimensional unitary conformal theory of identical operators with scaling dimensions less than one. With these assumptions in place we can entertain ourselves with some contour deformations as follows.

We begin by pulling the entire $+i\epsilon$ part of the integral in \eqref{ftcontourto1} upward in the complex $z$ plane. We first deform the contour smoothly until it lies just above the cut from $1$ to $+ \infty$, and then proceed by pulling it through the cut (from above) and continue to move it on the second sheet so that it now lies just below the negative real axis from $- \infty$ to $1$. This replaces $f(z)$ and $k_{-\a + 1/2}(z)$ with their continuation on the second sheet and changes the sign of the $i\e$ term. Mutatis mutandis we do the same for the $-i \epsilon$ part of the integral to find that:
\be
\ft(\a)  = \frac{1}{2\pi i} \int_{-\infty}^1 \frac{dz}{z^2}  \left( f^\circlearrowleft(z + i \e) k^\circlearrowleft_{- \a + 1/2}(z + i \e)  - f^\circlearrowright(z - i \e) k^\circlearrowright_{- \a + 1/2}(z - i \e) \right)\,,
\ee
where the circle indicates the direction of the continuation around $z = 1$. Next we can use
\be
\begin{split}
k^\circlearrowright_{- \a + 1/2}(z) &= k_{- \a + 1/2}(z) - 2 \pi i \, \a Q(\a)  \Psi_\a (z)\,,\\
k^\circlearrowleft_{- \a + 1/2}(z) &= k_{- \a + 1/2}(z) + 2 \pi i \, \a Q(\a)  \Psi_\a (z)\,,
\end{split}
\ee
and substitute $\Psi_\a(z) = \hf(Q(\a) k_{\a + 1/2}(z) + (\a \to -\a))$. We now leave everything that involves $k_{\a + 1/2}(z)$ on the secondary sheets whereas we move everything that involves $k_{-\a+1/2}(z)$ back to the first sheet. A little bit of reshuffling gives
\be
\begin{split}
\ft(\a) = \frac{\k_\a}{2} \int_{-\infty}^1 \frac{dz}{z^2}
\Big[ &f(z + i\e) k_{\a + 1/2}(z+i \e)(1 + i \cot(\pi \a))\\
&+f(z- i\e) k_{\a + 1/2}(z-i \e)(1 - i \cot(\pi \a))\\
&-f^\circlearrowright(z- i\e) k_{\a + 1/2}(z-i \e)\\
&-f^\circlearrowleft(z+ i\e) k_{\a + 1/2}(z+i \e)\Big]
\end{split}
\ee
with
\be
\k_\a \colonequals \frac{\G(\a + \hf)^4}{4\a\pi^2 \G(2\a)^2} = \frac{1}{\a \pi^2 Q(-\a)^2}\,.
\ee
For sufficiently large $\a$ the integrand should be finite as $z \to 0$ on all the sheets. In that case we can split the contour integrals up at $0$ into two parts that lie along the real $z$ axis. There are no branch cuts from $0$ to $1$ so in this bit the $i \epsilon$ prescription is unnecessary. For the bit from $-\infty$ to $0$ we will use
\be
k_h(z \pm i \e) = e^{\pm i \pi h} \ktil_h (z)\,, 
\ee
with\footnote{For later reference also note that, for $ w < 0$ and $ 0 < z < 1$, $\ktil_{h}(w) = k_h\left(\frac{w}{w-1}\right)$ and $k_{h}(z) = \ktil_h\left(\frac{z}{z-1}\right)$.
}
\be
\ktil_{h}(z) \colonequals (-z)^h {}_2 F_1(h,h,2h,z)\,.
\ee
This leads us to
\be \label{almostddisc}
\begin{split}
\ft(\a) &= \k_{\a} \int_0^1 \frac{dz}{z^2} k_{\a + 1/2}(z) \big[f(z) - \hf f^{\circlearrowleft}(z) - \hf f^{\circlearrowright}(z)\big]\\
&+ \frac{\k_{\a}}{2\cos(\pi(\a + 1/2))} \int_{-\infty}^0 \frac{dz}{z^2} \ktil_{\a + 1/2}(z) \big[f(z + i \e) + f(z - i \e)\big]\\
&- \frac{\k_{\a}}{2} \int_{-\infty}^0 \frac{dz}{z^2} \ktil_{\a + 1/2}(z) \big[f^\circlearrowright(z - i \e) e^{-i \pi(\a + 1/2)} + f^\circlearrowleft(z + i \e) e^{i \pi(\a + 1/2)}\big]\,.
\end{split}
\ee
Now, we are mostly interested in the OPE coefficients $\lambda_m$, and these are given by evaluating $\ft(\a)$ at $\a = p + m - 1/2$ with integer $m$. For these values of $\alpha$ we find
\be
\begin{split}
\lambda_m &= \Bigg\{ \k_{\a} \int_0^1 \frac{dz}{z^2} k_{\a + 1/2}(z) \big[f(z) - \hf f^{\circlearrowleft}(z) - \hf f^{\circlearrowright}(z)\big]\\
&+ (-1)^m \frac{\k_{\a}}{2\cos(\pi p)} \int_{-\infty}^0 \frac{dz}{z^2} \ktil_{\a + 1/2}(z) \big[f(z + i \e) + f(z - i \e)\big]\\
&- (-1)^m \frac{\k_{\a}}{2} \int_{-\infty}^0 \frac{dz}{z^2} \ktil_{\a + 1/2}(z) \big[f^\circlearrowright(z - i \e) e^{-i \pi p} + f^\circlearrowleft(z + i \e) e^{i \pi p}\big] \Bigg\}_{\a = p + m - 1/2}\,.
\end{split}
\ee
The first bit is the double discontinuity of \cite{Caron-Huot:2017vep} of $f(z)$ around 1. The second bit can also be written as a double discontinuity around minus infinity, but of a function $\widetilde{f(z)}$ which for $z < 0$ is defined as
\be
\widetilde{f(z)} \colonequals \frac{f(z+ i \e) + f(z- i\e)}{2 \cos(\pi p)}\,,
\ee
and for other values of $z$ via analytic continuation. Its block decomposition reads simply
\be
\widetilde{f(z)} = \sum_{m = 0}^\infty \lambda_m (-1)^m \ktil_{p + m}(z)\,.
\ee
Altogether we can then write that
\be \label{lambdamfromddisc}
\begin{split}
\lambda_m &= \ft^{(t)}(p + m -1/2) + (-1)^m \ft^{(u)}(p + m - 1/2)\\
\ft^{(t)}(\a) &\colonequals \k_{\a} \int_0^1 \frac{dz}{z^2} k_{\a + 1/2}(z) \text{dDisc}_1[ f(z) ]\\
\ft^{(u)}(\a) & \colonequals \k_{\a} \int_{-\infty}^0 \frac{dz}{z^2} \ktil_{\a + 1/2}(z) \text{dDisc}_\infty [\widetilde{f(z)}]\\
&= \k_{\a} \int_0^1 \frac{d z}{z^2} k_{\a + 1/2}(z) \text{dDisc}_1\left[ \reallywidetilde{f\left(z/(z-1)\right)}\right]\,.
\end{split}
\ee
This is of course reminiscent of the `Lorentzian inversion formula' of \cite{Caron-Huot:2017vep}, for which we have now derived a one-dimensional version in the case where $f(z)$ has an integer-spaced conformal block decomposition.

The somewhat odd manipulations we had to perform to arrive at $\ft^{(u)}(\a)$ are actually redundant in a two-variable function $f(z,\zb)$. In that case, whenever $z$ is continued to negative values in the above, we can also continue $\zb$ to negative values. This leads to extra phase factors and avoids the need to introduce $\widetilde{f(z)}$. We will discuss the two-variable case in more detail below.

\subsubsection{Examples}
We now have two `inversion formulas' that give the OPE coefficients of a function $f(z)$ with integer-spaced conformal block decomposition. To compare the different expressions it is instructive to consider the above examples again.

\begin{figure}
\begin{center}
\includegraphics[width=10cm]{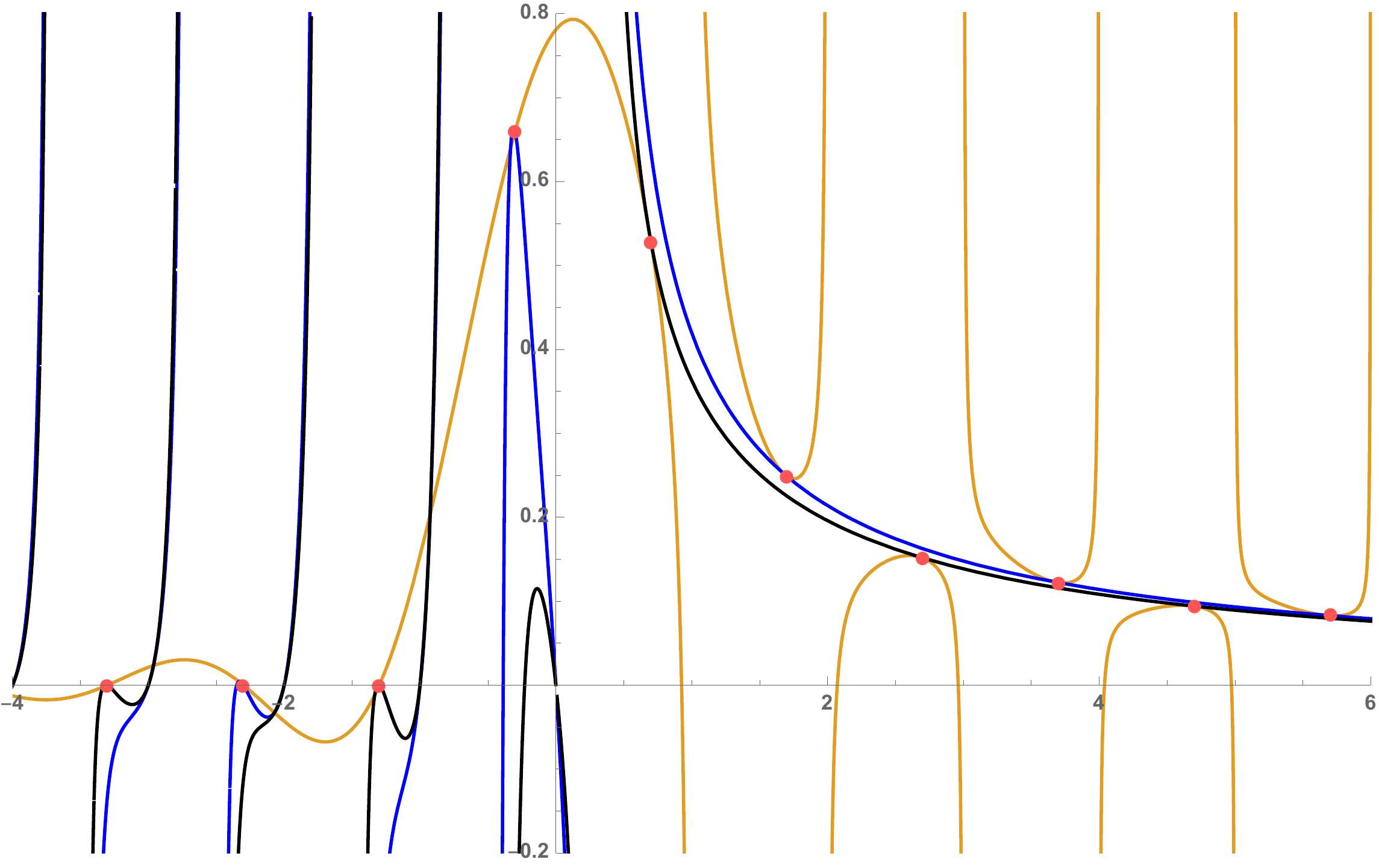}
\end{center}
\caption{\label{fig:ddiscequalsfcheck}A visualization of equation \eqref{ftequalsddisc} corresponding to the equivalence of the two inversion formulas for $z^p (1-z)^{-q}$ with $p = 0.2$ and $q = 0.3$. The orange curve is the left-hand side, so $\ft(\a)$, and the blue and black curves are the two expressions on the right-hand side, so $\ft^{(t)}(\a) + \ft^{(u)}(\a)$ and $\ft^{(t)}(\a) - \ft^{(u)}(\a)$. As indicated by the red dots, we find agreement for $\alpha = p + m -1/2$ where the correct $\lambda_m$ are reproduced. (For clarity of presentation we have divided all functions by $4^{-\a}$.)}
\end{figure}

Our first example is a sanity check of our derivations. We consider $z^p (1-z)^{-q}$. If we set
\be
I^{(s)}_{p,q}(\a) = \int_0^1 \frac{dz}{z^2}k_{\a + 1/2}(z) z^p (1-z)^{-q}\,,
\ee
and use that
\be
\begin{split}
\text{dDisc}_1 [ z^p (1-z)^{-q} ] &= 2 \sin^2(\pi q) z^p (1-z)^{-q}\,,\\
\text{dDisc}_\infty [ \reallywidetilde { z^p (1-z)^{-q} }] &= 2 \sin^2(\pi(p - q)) (-z)^p (1-z)^{-q}\,,
\end{split}
\ee
then the equivalence of the two inversion formulas should imply that
\be \label{ftequalsddisc}
I^{(C)}_{p,q}(p + m -1/2) = \left. \k_{\a} \left(2 \sin^2(\pi q) I^{(s)}_{p,q}(\a) + (-1)^m 2 \sin^2(p - q)  I^{(s)}_{p,p-q}(\a)\right)\right|_{\a = p + m - 1/2}\,.
\ee
This indeed works very well for any integer $m$ (including $m < 0$), as we illustrate in figure \ref{fig:ddiscequalsfcheck}.

Next we can consider the case where $f(z)$ is a single $s$-channel block, say $k_{\b + 1/2}(z)$. Then $\widetilde{f(z)} = \ktil_{\b + 1/2}(z)$ and from the equations given above it is not hard to deduce that both dDisc's are identically zero and therefore equation \eqref{lambdamfromddisc} trivially states that $\lambda_m = 0$ for all $m$. This is a pretty good score: compared to the exact answer, which is $\lambda_m = \delta_{m,0}$ (and $p = \beta + 1/2$), we correctly obtained all but one of the OPE coefficients! Notice that the missing OPE coefficient was correctly reproduced above by our starting point, which for the case at hand was $J^{(C)}(\a,\b)$ as plotted in figure \ref{fig:JCab}. It is therefore the preceding derivation that should be invalid for $m = 0$, and this is indeed the case: because of the well-known singularity in the `Regge' limit $z \to 0$ of $k^{\circlearrowleft}_{\b + 1/2}(z)$ and $k^{\circlearrowright}_{\b + 1/2}(z)$ we could not really push the contour on the secondary sheets all the way to 0 for $\alpha \leq \beta$, thereby invalidating the split of the integral into the $t$- and $u$-channel parts.

Finally we can consider conformal block expansions in the crossed channels. Let us first write the $t$-channel block decomposition as
\be
f(z) = \left(\frac{z}{1-z}\right)^h \sum_q \mu_q k_{h_q}(1-z)\,.
\ee
As we discussed around equation \eqref{ICpqtchannelblocks}, the original contour integral formula also commutes with the $t$-channel block decomposition and we can write
\be
\ft(\a) = \sum_s \mu_s K^{(C)}_h(\a, h_s - 1/2)\,.
\ee
On the other hand, for the dDisc formula we first of all have to realize that the integral in $\ft^{(u)}(\a)$ is entirely over a region where the $t$-channel block decomposition does not converge and we need to instead consider the `$u$-channel' conformal block decomposition of $\widetilde{f(z)}$. It reads:
\be
\widetilde{f(z)} = (-z)^h \sum_r \tilde \mu_r \ktil_{h_r}(1/z)\,,
\ee
and the corresponding dDisc inversion formula would read
\be \label{tudiscinversion}
\begin{split}
\lambda_m &\overset{!}{=} 2  \k_{\a} \left( \sum_{q} \mu_q \sin^2(\pi(h_q - h)) L^{(s)}_h(\a,h_q - 1/2) \right. \\ &\qquad \qquad \left. \left. + (-1)^m \sum_{r} \tilde \mu_r \sin^2(\pi(\tilde h_r - h)) L^{(s)}_h(\a,\tilde h_r - 1/2) \right)\right|_{\a = p + m -1/2}\,,
\end{split}
\ee
at least for those values of $m$ where the aforementioned subtleties related to the Regge limit do not come into play. Here we used the doubly split kernel
\be
L^{(s)}_h(\a,\b) \colonequals \int_0^1 \frac{dz}{z^2} \left(\frac{z}{1-z}\right)^h k_{\a + 1/2}(z) k_{\b + 1/2}(1-z)\,.
\ee
and the sine factors arise from taking the dDisc of the individual $t$- and $u$-channel blocks.

\begin{figure}
	\begin{center}
		\includegraphics[width=7.3cm]{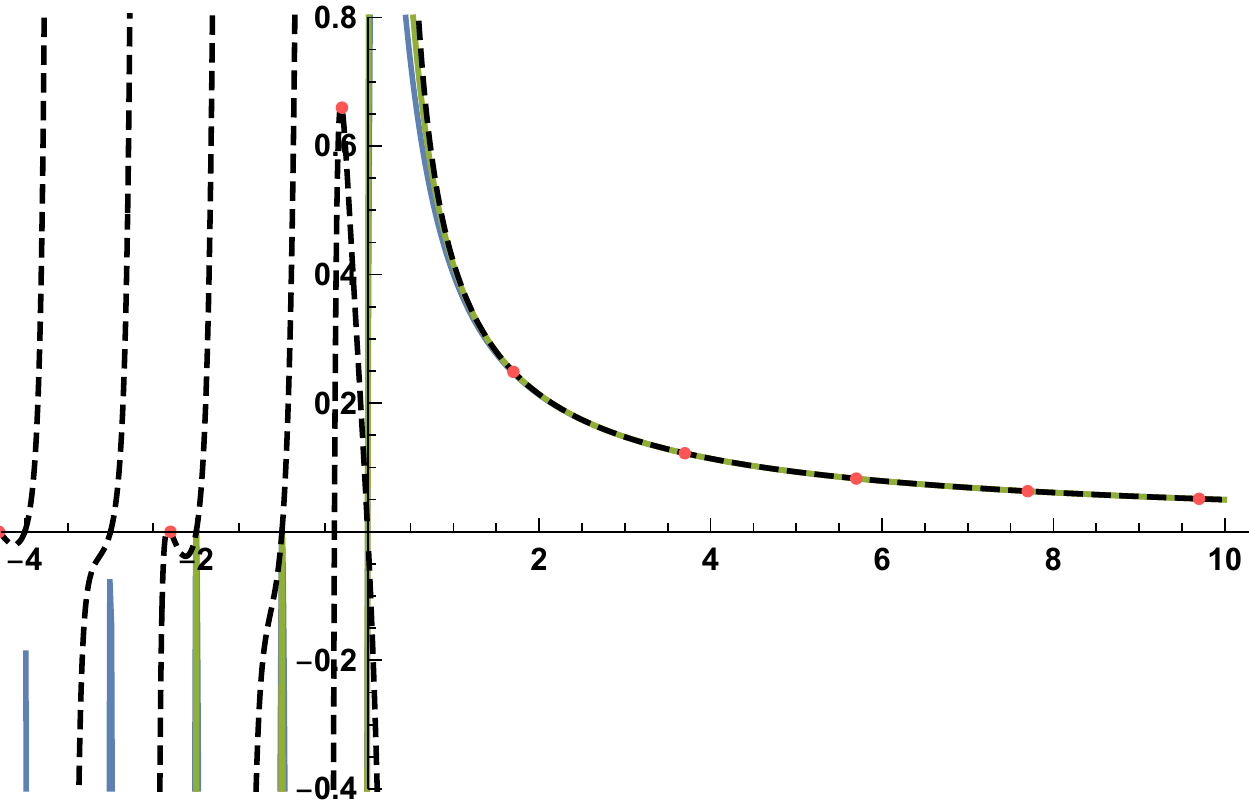}$\,$
		\includegraphics[width=7.3cm]{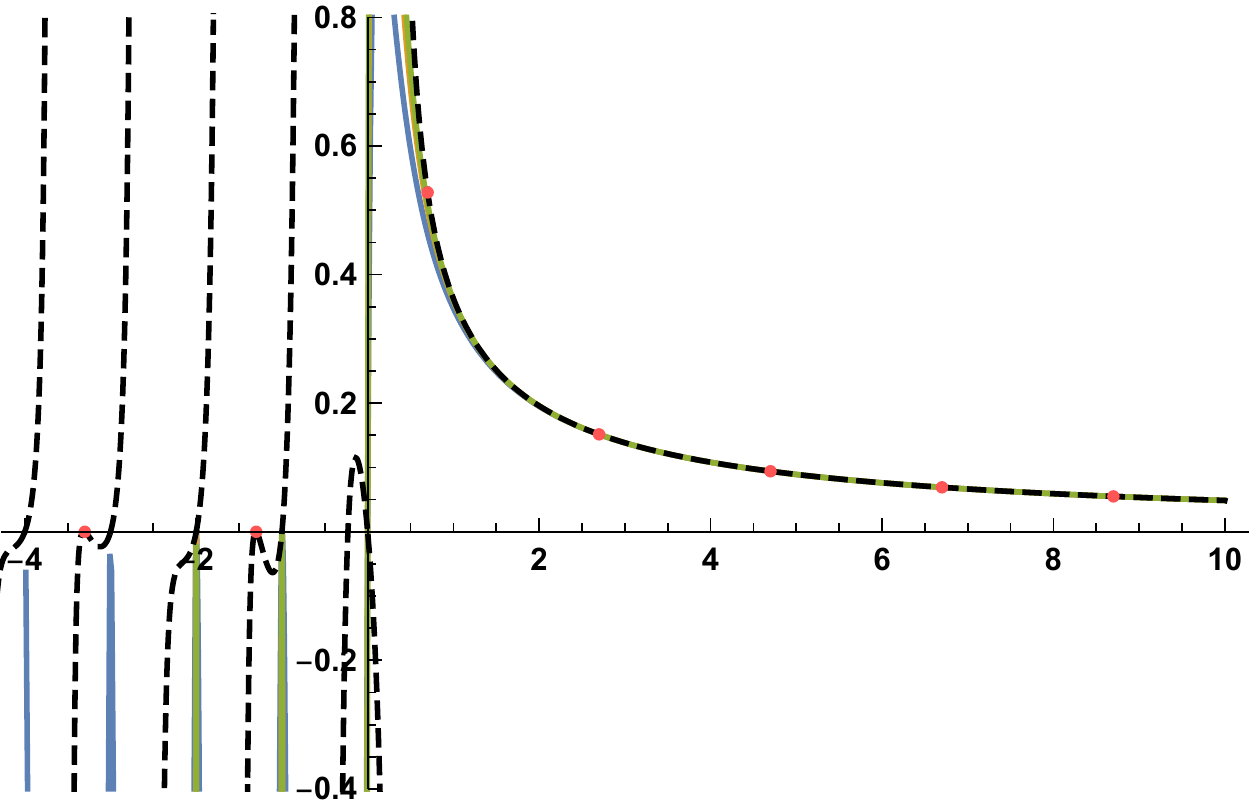}
		\caption{\label{fig:tuchannelD}Convergence of the $t$-channel block decomposition in the `dDisc inversion formula' \eqref{tudiscinversion} for $z^p(1-z)^{-q}$ with $p = 0.2$, $q=0.3$ and $h = 0.7$. We plot $\ft^{(t)}(\a) + \ft^{(u)}(\a)$ on the left and $\ft^{(t)}(\a) - \ft^{(u)}(\a) $ on the right. The dotted lines correspond to the exact expression, and the red dots are the $\lambda_m$ for $\a = p + m -1/2$. The solid lines correspond to approximations where we keep only 2 (blue, behind orange), 5 (orange) and 8 (green) terms in each of the cross-channel conformal block decompositions. (For clarity of presentation we have divided all functions by $4^{-\a}$.)}
	\end{center}
\end{figure}

Let us again consider the function $z^p (1-z)^{-q}$ as an example. The corresponding cross-channel decompositions are
\begin{align}
h_s &= h - q + s & \mu_q &= I^{(C)}_{h - q,h -p}(h-q - 1/2 + s) & s &\in \{0,1,2,\ldots\}\,,\\
\tilde h_r &= h - p + q + r & \tilde \mu_r &= I^{(C)}_{h - p + q, h-p}(h - p + q -1/2 + r) & r &\in \{0,1,2,\ldots\}\,.
\end{align}
In figure \ref{fig:tuchannelD} we show what happens if we plug these into 
\eqref{tudiscinversion} for $p = 0.2$, $q = 0.3$, $h = 0.7$, and keep only finitely many terms. Note that the exact result for the same values of $p$ and $q$ was already shown in figure \ref{fig:ddiscequalsfcheck}. It is most interesting to compare this cross-channel approximation with figure \ref{fig:tchanneldec} (these are the same OPE coefficients but the normalizations used in the plots are different). We see that the one-dimensional `dDisc inversion formula' gives an excellent approximation for all $m \geq 1$ and provides a smooth function through the OPE coefficients with even and odd $m$. (It is not entirely clear from the figure, but the sum over $t$-channel blocks is also monotonically increasing.) The approximation with the `contour inversion formula' in figure \ref{fig:tchanneldec}, by contrast, is not as good, not monotonic and not `smooth' in the sense that the interpolating function has poles on the real $\a$ axis. On the other hand, it is able to capture the $m = 0$ OPE coefficient (at $\a = -0.3$), which is completely missed by the dDisc inversion formula, and, less importantly perhaps, all the zeroes at negative $m$. In this sense the dDisc inversion formula sacrifices convergence for $m \leq 0$ in favor of excellent convergence for $m \geq 1$. 

Let us finally remark that the comparison above used twice the OPE data for the dDisc inversion formula since we kept 2, 5 and 8 blocks in \emph{both} the $t$-channel and the $u$-channel, compared to only $t$-channel blocks in the contour inversion formula. One may correct for this, but it does not meaningfully change the previous discussion about the merits of each inversion formula.

\subsection{Two variables}
\label{subsec:twovars}
It is now natural to define a transform of the form
\be
\frac{1}{2\pi i} \int_0^1 \frac{dz}{z^2} \int_C \frac{d\zb}{\zb^2}\Psi_\a (z) k_{-\ab + 1/2}(\zb) f(z,\zb)\,.
\ee
A single s-channel block would transform to
\be
k_{\b + 1/2}(z) k_{\bb + 1/2}(\zb) + (\b \leftrightarrow \bb) \qquad \to \qquad \frac{Q(-\b)}{2}\left(\frac{1}{\b - \a} + \frac{1}{\b+ \a}\right) J^{(C)}(\ab,\bb) + (\b \leftrightarrow \bb)\,.
\ee
Let us focus on the pole near $\a = \b$. If we set $\ab = j + \a$ and let $j$ be an integer then we find the behavior
\be
\frac{\delta_{j, J} Q(-\b)/2 }{\b - \a} \,,
\ee
with $J = \bb - \b$ the spin of the s-channel block that we are transforming. On the other hand, the residue of the shadow pole at $\a = - \b$ is not directly related to an OPE coefficient and therefore less useful. To get rid of the shadow poles we can once more split the density and introduce:
\be
\fd(\a,\ab) =  \frac{Q(-\a)}{2\pi i} \int_0^1 \frac{dz}{z^2} \int_C \frac{d\zb}{\zb^2}k_{-\a+1/2} (z) k_{-\ab + 1/2}(\zb) f(z,\zb)\,,
\ee
which removes the pole at $\a = -\b$.

Now, for a general CFT four-point function (with a decomposition into integer spin blocks) it is tempting to believe the fundamental meromorphicity property persists: we claim that for integer values of $\ab \pm \a$, the transformed function has poles precisely when $\a$ equals the twist of an $s$-channel block with residues given in terms of the corresponding OPE coefficients. In simple examples this property directly follows from the general properties of the one-dimensional alpha space transforms in combination with the manipulations of the previous sections, for example in all cases where $f(z,\zb)$ factorizes like $f_1(z) f_2(\zb)$ or equals a finite sum of such products. But whether it is true more generally remains a conjecture at this point.

The density $\fd(\a,\ab)$ appears to have the right analyticity structure to make contact with the density that can be obtained from the Euclidean and Lorentzian inversion formulae of \cite{Dobrev:1977qv} and \cite{Caron-Huot:2017vep}. Let us make this connection more precise. By virtue of the integrality of the spin, for every term of the form $z^{h}f_h(\zb)$ in the small $z$ expansion of $f(z,\zb)$ the corresponding $f_h(\zb)$ will have an integer-spaced conformal block decomposition. This means that we are allowed to perform the contour pulling exercise of the previous section for the $\bar z$ integral and replace this integral with \eqref{almostddisc}. For the parts contributing to $\ft^{(t)}(\ab)$ in \eqref{lambdamfromddisc} we can leave the $z$ integral in place, whereas for the parts contributing to $\ft^{(u)}(\ab)$ we can rotate the $z$ integral into the lower or upper half plane so it is parallel to the negative real axis. This latter rotation introduces an analytic part from the contribution from 1 to $\infty$, which does not contribute dynamical poles in $\a$, and also some phase factors which (if we rotate in the correct sense) precisely ensure that the single-variable dDisc$_{\infty}$ in \eqref{lambdamfromddisc} becomes the original two-variable dDisc$_{\infty}$ of \cite{Caron-Huot:2017vep}. Finally, if we shadow-symmetrize the density as in
\be
\fd(\a,\ab) + \fd(-\ab,-\a)\,,
\ee
then we find ourselves integrating this dDisc against a `block with dimension and spin interchanged', exactly as in \cite{Caron-Huot:2017vep}. So this explains how alpha space in two dimensions can be related to the Lorentzian (and therefore also Euclidean) inversion formulas.\footnote{An important ingredient for the other inversion formulas to work is single-valuedness of the Euclidean correlator. Even though our prescription started entirely in the Lorentzian square, this property did enter our derivation via the demand for an $s$-channel conformal block decomposition with integer spins.} 

\subsection{\texorpdfstring{Sum over $t$-channel blocks}{Sum over t-channel blocks}}
\label{subsec:tchannelsumanalytic}
\begin{figure}
\begin{center}
\includegraphics[width=7cm]{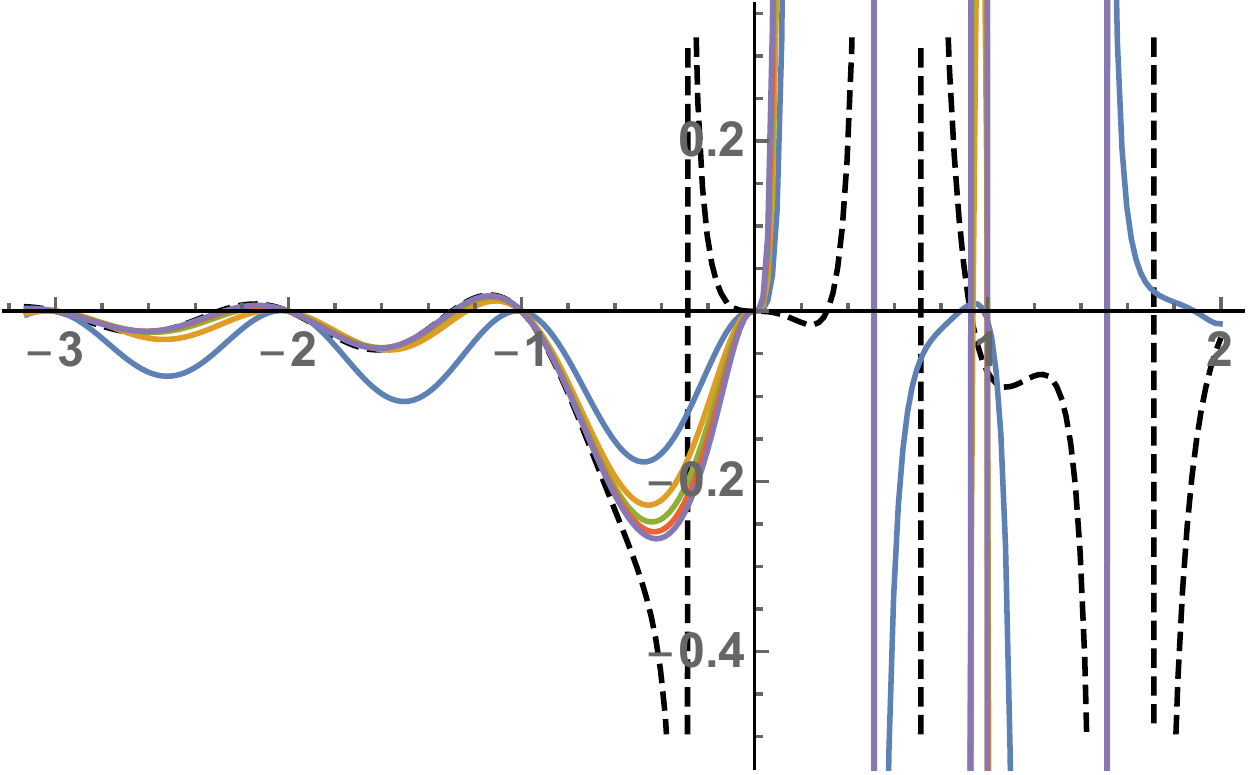}$\quad$\includegraphics[width=7cm]{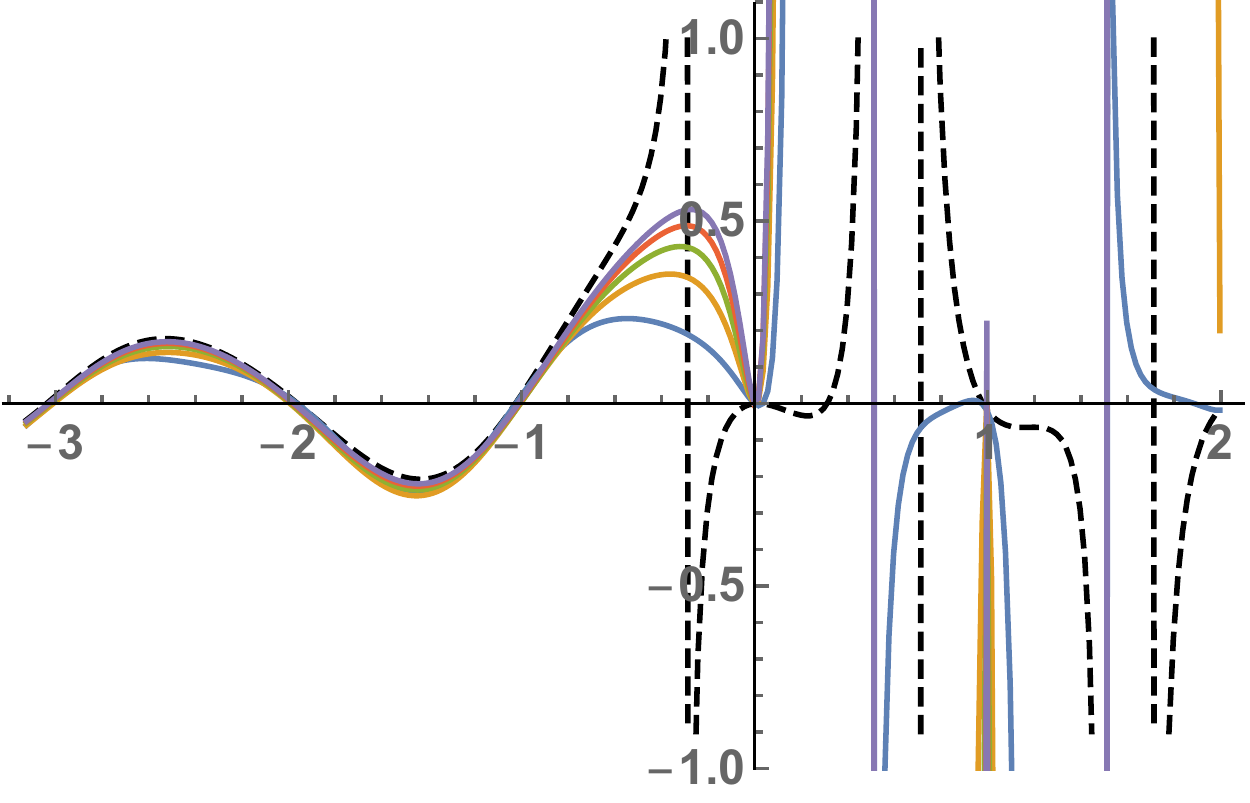}
\end{center}
\caption{\label{fig:fdtchannelblocks}In black, dashed: the transform $\fd(\a,\a + j)$ for the function given in equation \eqref{zzbpq} as a function of $\a$ with $j = 0$ (left) and $j = 6$ (right). We have divided by a normalization factor $Q(\a)\G(\pm(\a + j))4^{-\a-j}$ which in particular got rid of all the kinematic poles. Colored lines: the approximation to this density by transforming a finite sum of the first few $t$-channel blocks, ranging from a single block (blue line) to 45 blocks (purple line). As expected, we observe good convergence for negative $\a$. Since we chose $\D_\phi = 0.7$ the individual $t$-channel blocks only produce poles at $\a = 0.2$ and so we can only approximately observe the actual pole at $\a = - 0.3$.}
\end{figure}

We can once more consider the two-variable function $f(z,\zb)$ as a sum over $t$-channel blocks and try to swap the integrals over $z$ and $\bar z$ with this sum, in the vein of equations \eqref{KsplitKsplitsum} or \eqref{KdsplitKsplitsum}. For our more analytic density we would have
\be \label{KdsplitKCsplitsum}
\fd(\a,\ab) \overset{?}{=} \sum_{k} R_k \left( K_\text{dsplit}(-\a,\b_k | \D_\phi) K^{(C)}_{\D_\phi}(\ab,\bb_k) + (\b_k \leftrightarrow \bb_k) \right)\,.
\ee
As follows from the previous analyses, we should be able to commute the sum with the above integrals when $\a$ is sufficiently negative. In fact, it appears that $\ab$ can kept finite, see for example the convergence everywhere in figure \ref{fig:tchanneldec}. Let us work this out in the two-variable version of the simple example we considered before, that is we decompose
\be \label{zzbpq}
\frac{(z\zb)^p}{[(1-z)(1-\zb)]^q}\,,
\ee
with $p = 0.2$ and $q = 0.3$ into $t$-channel blocks with $\Delta_\phi = 0.7$. The formulas are entirely analogous to the discussion around figure \ref{fig:tchanneldec} and we will not repeat them here. For spins 0 and 6 we plot the results in figure \ref{fig:fdtchannelblocks}, which clearly exhibits the good convergence properties for negative $\a$.

\subsubsection*{Discussion}

An interesting question is how much information about the $s$-channel OPE data can be obtained from a finite sum of $t$-channel blocks. Early encouraging results \cite{Alday:2015ewa} showed how we can obtain surprisingly accurate estimates of at least the leading Regge trajectory from only a few $t$-channel blocks in the four-point function of the $\sigma$ operator in the three-dimensional Ising CFT. This analysis was substantially improved with a more systematic analysis of multiple correlation functions in \cite{Simmons-Duffin:2016wlq}. The Lorentzian inversion formula of \cite{Caron-Huot:2017vep} not only provides a more in-depth understanding of this phenomenon by appealing to analyticity in spin; in practice its use also somewhat improves \cite{Albayrak:2019gnz} the estimates of the spectrum at low spins.

More generally, sums like \eqref{KdsplitKCsplitsum} and similar expressions discussed either in this paper or elsewhere are often used to obtain an approximation of the $s$-channel spectrum from a finite number of $t$-channel blocks (or Regge trajectories). As we have seen, these approximations appear to be reliable at large $J$ (or at large $\Delta$, see \cite{Mukhametzhanov:2018zja}). For finite $J$ and $\Delta$ there is no small parameter and the error is not under control but in several cases the estimates turn out to be remarkably close to numerical bootstrap results (see for example the above references, but also \cite{Cornagliotto:2017snu} for a four-dimensional supersymmetric example).

With this in mind it might be interesting to take a closer look at our formulas. First of all, in equations \eqref{KsplitKsplitsum} and \eqref{KdsplitKsplitsum} we find that each $t$-channel block contributes double poles in \emph{both} $\a$ and $\ab$. This reflects the fact that a single $t$-channel block \emph{cannot} be expressed as a sum of $s$-channel blocks or $\Delta$-derivatives of $s$-channel blocks: to reproduce the $\log(z) \log(\bar z)$ one would have to take both a $\Delta$-derivative and a $J$-derivative of an $s$-channel block. In the lightcone bootstrap we circumvented this issue by considering the disc$_\ab$, getting rid of one of the double poles, and interpreting the other as a small anomalous dimension at large $\ab$. Formulas like \eqref{KdsplitKCsplitsum} but also the Lorentzian inversion formula handle the issue differently: they presuppose that $f(z,\zb)$ can be decomposed into integer spin $s$-channel blocks, and if one plugs in a finite number of $t$-channel blocks then the transform itself ensures that the result is cast into a form that can be interpreted as a sum over $s$-channel blocks or $\Delta$-derivatives thereof. Of course one loses information in the process: indeed, for these transforms one does not obtain a single $t$-channel block if one reverts its density back to position space.

Finally, the main difference between \eqref{KdsplitKCsplitsum} and the Lorentzian inversion formula is of course analyticity in spin. As noted already in the one-dimensional case, \eqref{KdsplitKCsplitsum} appears to converge to the right answer for all spins $J \geq 0$, albeit much more slowly than in the Lorentzian inversion formula. It would be interesting to consider more realistic correlation functions, for example those in the three-dimensional Ising model decomposed into two-dimensional blocks \cite{Hogervorst:2016hal}, and see how well \eqref{KdsplitKCsplitsum} compares to the Lorentzian inversion formula both at spin 0 and also at higher spins. This might in particular provide a nice alternative tool to analyze the deep Euclidean limit as in \cite{Mukhametzhanov:2018zja}. We leave more investigations to future work.

\section*{Acknowledgments}
We would like to thank Simon Caron-Huot, Matthijs Hogervorst, Mikhail Isachenkov, Madelena Lemos, Marco Menieri, Slava Rychkov and Volker Schomerus for their insightful comments. We greatly benefited from discussions during the 2019 Pollica summer workshop and during a visit to IHES, and we are therefore grateful for the hospitality provided. BvR is supported in part by the STFC under consolidated grant ST/P000371/1 and a grant from the Simons Foundation (\#488659).

\appendix 

\section{Further alpha space densities}
\label{app:weird}
In this appendix we give a few examples of `exotic' alpha space densities for various functions and distributions. We will restrict ourselves to one-dimensional alpha space.

\subsection{Polynomials in alpha space}
Consider an alpha space density that is a polynomial in $\a^2$. Much like in ordinary momentum space, the corresponding position-space expression turns out to be a derivative of a delta function. To see this it suffices to consider the density $\fh(\a) = 1$ since all other polynomials can be obtained by finite action of the Casimir operator. There are many ways to see that this density gives a delta function centered at one in position space. For example, we can consider
\be
f(z) = \frac{z^p}{p} \qquad \Leftrightarrow \qquad \fh(\a) = \frac{\G( p - 1/2 \pm \a)}{p\G(p)^2}\,.
\ee
Now take the large $p$ limit of both sides we get
\be
f(z) = \lim_{p \to \infty} \frac{z^p}{p} = \delta(z - 1) \qquad \Leftrightarrow \qquad \fh(\a) = 1\,.
\ee
Notice that the support of the delta function is understood to lie entirely within the range of a $z$-integral from $0$ to $1$, as is stated more precisely by its definition as a limit.

A more elegant derivation is the following. We start from:
\begin{equation}
\log(1-z)\qquad \Leftrightarrow \qquad \frac{1}{\a^2-\frac{1}{4}}
\end{equation}
and use Parseval's formula to say that
\begin{align}
\vev{D_z \log(1-z),f(z)}&=\int_0^1 \frac{dz}{z^2}D_z(\log(1-z))f(z)\\
&=\int\left[d\a\right]\frac{2\hat{f}(\a)}{Q(\a)Q(-\a)}\left(\frac{\a^2-\frac{1}{4}}{\a^2-\frac{1}{4}}\right)\nonumber\\
&=\int\left[d\a\right]\frac{2\Psi_\a(1)\hat{f}(\a)}{Q(\a)Q(-\a)}\nonumber\\
&=f(1)\nonumber
\end{align}
where we have used the fact that
\begin{equation}
D_z f(z)\qquad \Leftrightarrow \qquad \left(\a^2-\frac{1}{4}\right)\hat{f}(\a)\,.
\end{equation}
Therefore, an alpha space density of 1 corresponds to a position space function of $D_z\log(1-z)$ which is equal to $z^2\,\d(1-z)$.

\subsection{The constant function}
Next we consider the alpha space transform of $f(z) = 1$. We could consider the $p \to 0$ limit of the above expression but this is a little nasty because of the coalescence of three poles at $\a = \pm 1/2$. It is simpler to start from
\be
f(z) = k_h(z) \qquad \Leftrightarrow \qquad \fh(\a) = \frac{-\G(2h)/\G(h)^2}{\a^2 - (h- 1/2)^2} \equalscolon \fh_h(\a)\,,
\ee
and take the limit $h \to 0$ for which the left-hand side clearly becomes 1. On the right-hand side we find
\be \label{limh0fhh}
\lim_{h \to 0} \fh_h(\a) = 
\begin{cases}
1/2 & \text{if } \a = \pm 1/2\,,\\
0 & \text{otherwise.} 
\end{cases}
\ee
Strange as it may seem, we claim that the right-hand side is our best possible definition of the alpha space transform of the constant function. As a consistency check, notice that the action of the Casimir differential operator on the constant function is identically zero. In alpha space this means that the alpha space transform of the constant function, when multiplied with $\a^2 - 1/4$, must vanish identically as well and this clearly is the case for the given function.

More precisely, the $z$-integral defining the alpha space transform of $k_h(z)$ does not converge if $\Re(h) \leq 1/2$, and in the inverse transform this issue shows up because we have to deform the $\alpha$ contour of integration away from the imaginary axis. Precisely at $h = 0$ the contour then gets pinched between the poles of $\fh_h(\a)$ at $\a = \pm (h - 1/2)$ and poles at $\a = \pm 1/2$ arising from the measure $N(\a)^{-1}$. In order to avoid this pinching we can push the contour \emph{also} beyond the poles at $\a = \pm 1/2$ and subtract extra factor proportional to the residues there. Calling this new contour $\mathcal C$ we get
\be
k_h(z) = \fh_h(1/2) + \fh_h(-1/2) + \int_{\mathcal C} [d\a] \frac{1}{N(\a)} \Psi_\a(z) \fh_h(\a)\,.
\ee
Since $\mathcal C$ does not cross the points where $\a = \pm 1/2$, all is well if we send $h \to 0$ and use the right-hand side of \eqref{limh0fhh}.

Fundamentally, for functions that are too singular near $z = 0$ to be square integrable it is important to realize that the alpha space contour has to be deformed away from the imaginary axis \cite{Hogervorst:2017sfd}. Any alpha space transform like \eqref{limh0fhh} therefore must be supplemented with a prescription of how the integration contour is supposed to lie.

\subsection{Logarithms}
Since simple poles in alpha space map to conformal blocks, it is sensible to also appraise higher-order poles. These singularities roughly correspond to derivatives of blocks. As an example, consider the integral
\begin{equation}
-\frac{Q(-\b)}{2}\int\left[d\a\right]\frac{2\Psi_\a(z)}{Q(\a)Q(-\a)}\left(\frac{1}{(\a-\b)^2}+\frac{1}{(\a+\b)^2}\right)\,.
\end{equation}
Once again, the symmetry of the integrand allows for the substitution $\Psi_\a(z)\rightarrow Q(\a)k_{\a+1/2}(z)$. After this split, the contour can be closed to the right to pick up the double pole and give
\begin{equation}
Q(-\b)\frac{\partial}{\partial\b}\left(\frac{k_{\b+1/2}(z)}{Q(-\b)}\right)=\frac{\partial k_{\b+1/2}(z)}{\partial\b}+\tilde{\psi}(\b)k_{\b+1/2}(z)\,,
\end{equation}
where the function $\tilde{\psi}(\b)$ is defined to be
\begin{equation}\label{TildePsi}
\tilde{\psi}(\b)\equiv -\frac{d\log Q(-\b)}{d\b}=2 \psi\left(\frac{1}{2}+\b\right)-2\psi(2\b)\,,
\end{equation}
and $\psi(\b)$ is the digamma function
\begin{equation}
\psi(\b)\equiv \frac{d\log\G(\b)}{d\b}\,.
\end{equation}
Putting this all together, the first derivative of a block maps to alpha space like
\begin{equation}\label{D1DerivativeMap}
\frac{\partial k_{\b+1/2}(z)}{\partial\b}\qquad \Leftrightarrow \qquad -\frac{Q(-\b)}{2(\a-\b)}\left(\frac{1}{\a-\b}-\tilde{\psi}(\b)\right)+(\a\rightarrow -\a)\,.
\end{equation}
Such block derivatives produce $\log(z)$ terms and so the alpha space density for the logarithm deserves some attention.
\begin{figure}
	\centering
	\includegraphics[width=0.55\linewidth]{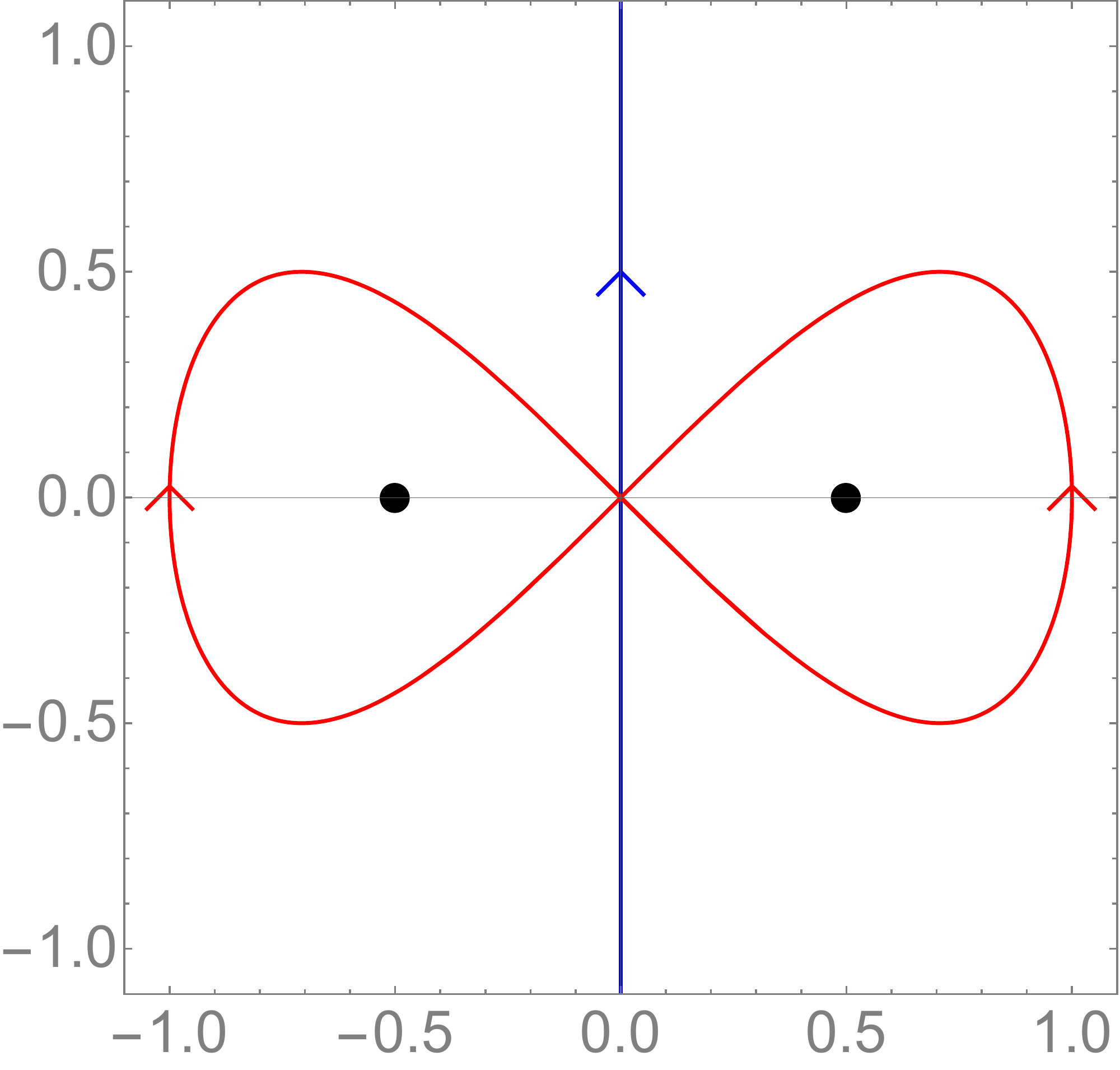}
	\caption[Contours for the Logarithm]{Plotted are the contours for the integral \cref{LogContourIntegral}. The blue contour gives $\log(1-z)$ whereas the figure-of-eight red one gives $2(1-\log(z))$. The double poles at $\a=\pm 1/2$ are marked by black dots.}
	\label{fig:LogContours}
\end{figure}
Consider the integral
\begin{equation}\label{LogContourIntegral}
\int\frac{d\a}{2\pi i}\frac{2\Psi_\a(z)}{Q(\a) Q(-\a)(\a-\hf)(\a+\hf)}
\end{equation}
over the two contours plotted in \cref{fig:LogContours}. The blue contour is the standard alpha space one. Therefore, $\Psi_\a$ can be replaced by $Q(\a) k_{\a+1/2}$ under the integral and the contour closed in the right half plane to give the single block
\begin{equation}
-\frac{2k_1(z)}{Q(-\frac{1}{2})}=-k_1(z)=\log(1-z)\,.
\end{equation}
However, two double poles are enclosed by the red contour because of the $Q$ factors and so the integral is
\begin{equation}
-4\lim_{\a\rightarrow -1/2}\frac{\partial}{\partial \a}\left(\frac{\Psi_\a(z)(\a+\hf)}{Q(\a) Q(-\a)(\a-\hf)}\right)=2(1-\log(z))\,.
\end{equation}
We note that in terms of the derivative of a block, the logarithm is
\begin{equation}
\log z=\left.\frac{\partial k_{\a+1/2}(z)}{\partial \a}\right|_{\a=-1/2}\,.
\end{equation}
The red contour may appear to be peculiar, but it arises naturally via the following argument. The derivative of $z^p$ with respect to $p$ is $z^p\log(z)$ and so a regularised alpha space density for the logarithm can be defined through
\begin{equation}
z^p\log(z)\qquad \Leftrightarrow \qquad\frac{\partial}{\partial p}\left( \frac{\G(p-\frac{1}{2}\pm\a)}{\G^2(p)}\right)\,,
\end{equation}
where we have used the density for $z^p$ as given above. We can get higher powers of the logarithm by taking further $p$ derivatives. The standard alpha space contour for this density is pinched at $\a=\pm \frac{1}{2}$ as $p\rightarrow 0$ and so it makes sense to pull the contour through $\a=\pm (\frac{1}{2}-p)$ with $p$ small, picking up the residues at these points and leaving an integral along the imaginary axis. The contribution of this integral vanishes as $p\rightarrow 0$ because the integrand vanishes, and so it can be dropped. The contribution of the residues as $p\rightarrow 0$ amounts to
\begin{equation}
\left.\frac{\partial \Psi_\a(z)}{\partial \a}\right|_{\a=-1/2}=\log(z)\,,
\end{equation}
and we see that everything is consistent. The two small contours around $\a=\pm (\frac{1}{2}-p)$ giving rise to these residues can be joined up to give a figure-of-eight path analogous to the red contour in \cref{fig:LogContours}.

Similar alpha space densities can be defined by substituting the alpha space transformation of $z^p$ into various Taylor series expansions. Two interesting examples are
\begin{equation}	\log(z)\log(1-z)\qquad \Leftrightarrow \qquad \frac{1}{(\hf+\a)(\hf-\a)}\left(2\,\G\!\left(\hf\pm\a\right)-\frac{1}{(\hf+\a)(\hf-\a)}\right)
\end{equation}
and the polylogarithm
\begin{equation}
\text{Li}_s(z)\qquad \Leftrightarrow \qquad\sum_{p=1}^\infty\frac{\G(p-\frac{1}{2}\pm\a)}{p^s\G^2(p)}=\G\left(\hf\pm\a\right)\!\,_{s+1}F_s\left(\begin{matrix}
\hf+\a,\hf-\a,1,...,1\\2,...,2
\end{matrix};1\right)
\end{equation}
for $s\in\mathbb{Z}^{\geq 1}$. Notice that, as always, the residues of the alpha space density immediately provide the spectrum and coefficients in the conformal block decomposition of the given functions.

\section{Proving a remarkable exact identity}
\label{app:DSDformula}
The author of \cite{Simmons-Duffin:2016wlq} procured a formula relating a Casimir singular term to a sum over conformal blocks, up to a sum over regular terms. In the remainder of this section, we re-derive their expression using the alpha space formalism.

A sum over derivatives of some function $f$ can be represented like
\begin{equation}
\sum_{k=0}^\infty \frac{d f(k)}{dk}=-\int\frac{ds}{2\pi i}\left(\frac{\pi}{\sin(\pi s)}\right)^2 f(-s)
\end{equation}
for a suitable contour. As such, the sum
\begin{equation}\label{DSDsum}
\left(\frac{1-z}{z}\right)^{p}+\frac{\G(\b-\hf-p)}{\G^2(-p)\G(\b+\hf+p)}\sum_{k=0}^\infty \frac{\partial }{\partial k}\left(\frac{\G(\b+\hf+k)}{(k!)^2(k-p)\G(\b-\hf-k)}\left(\frac{1-z}{z}\right)^k\right)\,,
\end{equation}
which is a Casimir singular term plus an infinite sum over regular terms, is equal to
\begin{equation}
\left(\frac{1-z}{z}\right)^{p}+\frac{\G(\b-\hf-p)}{\G^2(-p)\G(\b+\hf+p)}\int\frac{ds}{2\pi i}\frac{\G^2(s)\G(\b+\hf-s)}{(s+p)\G(\b-\hf+s)}\left(\frac{1-z}{z}\right)^{-s}\,,
\end{equation}
where the contour encircles the (double) poles at $s=-\mathbb{N}$ in a counter-clockwise manner.

Pulling this contour up to lie along the imaginary axis picks up the pole at $s=-p$, which contributes
\begin{equation}
-\left(\frac{1-z}{z}\right)^{p}
\end{equation}
to cancel against the $\left(\frac{1-z}{z}\right)^{p}$ term already present.

The remaining integral can then be taken to alpha space by transforming the $z$-dependent factor under the integral according to 
\begin{equation}
\left(\frac{1-z}{z}\right)^{-s}\mapsto \frac{\G(1-s)\G(s-\hf\pm\a)}{\G(s)\G(\hf\pm\a)}
\end{equation}
to give
\begin{equation}
\frac{\G(\b-\hf-p)}{\G^2(-p)\G(\b+\hf+p)\G(\hf\pm\a)}\int\left[ds\right]\frac{\G(s)\G(1-s)\G(\b+\hf-s)\G(s-\hf\pm\a)}{(s+p)\G(\b-\hf+s)}\,.
\end{equation}

This integral produces poles in the $\a$ plane because the $s$ contour separating left- and right-running semi-infinite sequences of poles is pinched whenever $\a=\pm(\b+\mathbb{N})$. As usual, these poles encode a block decomposition. The residues at the pinch-points prove that \cref{DSDsum} is equal to
\begin{equation}
\frac{1}{\G^2(-p)}\sum_{n=0}^\infty\frac{\G^2(\b+\hf+n)\G(\b-\hf-p+n)}{\G(2(\b+n))\G(\b+\frac{3}{2}+p+n)}k_{\b+n + 1/2}(z)\,,
\end{equation}
which reduces to the block decomposition of $\left(\frac{1-z}{z}\right)^{p}$ when $\b$ is tuned to $-p-\hf$.

The significance of this equality is that the residue of the pole at $\a=\b+n$ in alpha space is
\begin{equation}
-\left(\frac{Q(-\b-n)}{2}\right)\frac{\G^2(\b+\hf+n)\G(\b-\hf-p+n)}{\G^2(-p)\G(2(\b+n))\G(\b+\frac{3}{2}+p+n)}\,,
\end{equation}
which is asymptotically equal to
\begin{equation}
-\frac{n^{-2(1+p)}}{\G^2(-p)}\left(1+O\left(\frac{1}{n}\right)\right)\,.
\end{equation}
That is, the leading asymptotic behavior is independent of $\b$ and is exactly equal to the leading large $n$ asymptotic behavior of
\begin{equation}
-\frac{\G(-2p-1+n)}{n!\,\G^2(-p)}\,,
\end{equation}
which is the residue of the alpha space density for $\left(\frac{1-z}{z}\right)^p$ at $\a=-p-\hf+n$. In conclusion, the addition of the Casimir regular terms has shifted the locations of the blocks and their OPE coefficients without changing the leading asymptotics. A more complete discussion can be found in \cite{Simmons-Duffin:2016wlq}.

\section{Split kernel}
\label{app:splitkernel}
The split kernel can be written as
\be
K_\text{split}(\a,\b|\D_\phi) = \frac{2}{Q(-\b)} \int_0^1 \frac{dz}{z^2} \left(\frac{z}{1-z}\right)^{\D_\phi}\Psi_\a(z) k_{\b + 1/2}(1-z)\,.
\ee
It is understood to be defined through analytic continuation when its parameters are outside the region of convergence of the integral.  Using the Mellin-Barnes representation for $\Psi_\a(z)$ and the useful formula \eqref{usefulformula}, we immediately find that it can also be written as
\be
K_\text{split}(\a,\b|\D_\phi) = \frac{2 \b}{\G(\hf \pm \a)} \int [ds] \frac{\G(\hf + s \pm \a) \G(\b + s - \D_\phi+\frac{3}{2})\G(-s) \G(\D_\phi-s-1)^2}{\G(1 + s) \G(\b - s + \D_\phi - \hf)}\,.
\ee
In more detail, a representation of the split kernel is\footnote{For four-point functions with unequal external dimensions there exists a similar representation of the split kernel in terms of Wilson functions; it can be found in \cite{Dansthesis}.}
\begin{align}
K_\text{split}(\a,\b|\D_\phi)=\,&2\b\G(\D_\phi)\G(1-\D_\phi)\G\!\left(\D_\phi-\frac{1}{2}\pm\a\right)\!\G\!\left(\frac{1}{2}\pm\b\right) \times\\
&\Bigg(\Bigg.\frac{\G(\D_\phi-\frac{1}{2}\pm\a)W_\a(\b;\frac{1}{2},\frac{1}{2},\D_\phi-\frac{1}{2},\D_\phi-\frac{1}{2})}{\G(\frac{1}{2}\pm\a)}\,-\nonumber\\
&\phantom{\Bigg(\Bigg.}\frac{\G(\frac{3}{2}-\D_\phi+\b)W_\a(\b;\frac{1}{2},\frac{1}{2},\D_\phi-\frac{1}{2},\frac{3}{2}-\D_\phi)}{\G(\D_\phi-\frac{1}{2}+\b)}\Bigg.\Bigg)\,,\nonumber
\end{align}
where
\begin{equation}\label{WilsonDef}
W_\a(\b;a,b,c,d)\equiv\frac{\G(d-a)\,_4F_3\left(\begin{matrix}
	\tilde{a}-\a,\tilde{a}+\a,a-\b,a+\b\\
	a+b,a+c,a-d+1
	\end{matrix};1\right)}{\G(a+b)\G(a+c)\G(\tilde{d}\pm\a)\G(d\pm\b)}+(a\leftrightarrow d)
\end{equation}
is called the Wilson function\footnote{Another useful representation is in terms of a single well-poised $_7F_6$:
	\begin{equation*}
	W_\a(\b;a,b,c,d)=\frac{\G(2a+\tilde{d}-\a)\,_7F_6\left(
		\begin{matrix}
		\tilde{a}-\a,\tilde{b}-\a,\tilde{c}-\a,2a+\tilde{d}-1-\a,\frac{2a+\tilde{d}+1-\a}{2},a-\b,a+\b\\
		a+b,a+c,a+d,\frac{2a+\tilde{d}-1-\a}{2},\frac{a+b+c+d}{2}-\a-\b,\frac{a+b+c+d}{2}-\a+\b\end{matrix}\,;1\right)}{\G(a+b)\G(a+c)\G(a+d)\G(\tilde{d}+\a)\G(\frac{a+b+c+d}{2}-\a\pm\b)}
	\end{equation*}}. The dual variables in this definition are
\begin{equation}
\begin{pmatrix}
\tilde{a}\\\tilde{b}\\\tilde{c}\\\tilde{d}
\end{pmatrix}\equiv\frac{a+b+c+d}{2}-	\begin{pmatrix}
d\\c\\b\\a
\end{pmatrix}
\end{equation}
such that
\begin{equation}
W_\a(\b;a,b,c,d)=W_\b(\a;\tilde{a},\tilde{b},\tilde{c},\tilde{d})\,.
\end{equation}
The (tilded) split kernel has double poles when $\a$ approaches $h - 1/2 + n$ for non-negative integer $n$:
\be
\tilde K_\text{split}(\ab,\bb|\D_\phi) = \frac{M^{(K)}_n(\bb|\D_\phi)}{(\ab - \D_\phi + \hf - n)^2} + \frac{N^{(K)}_n(\bb|\D_\phi)}{\ab - \D_\phi + \hf - n} + \text{regular,}
\ee
with
\be
\label{mninsplitkernel}
\begin{split}
M^{(K)}_n(\b|h) &= \frac{\sin(\pi (h + n))}{\sin(\pi h)} \frac{\G(2\b + 1)\G(2h-1 + n)^2 \G(\hf - \b)}{\Gamma(n+1)^2 \G(\hf+\b)} W_{h -1/2 +n}\left(\b;\frac{1}{2},\frac{1}{2},h-\frac{1}{2},h-\frac{1}{2}\right)\\
N^{(K)}_n(\b|h) &=  \frac{(-1)^n  \pi \G(2\b+1)\G(2h-1+n)\G(\hf-\b) \G(\frac{3}{2}-h+\b)}{\sin(\pi h)\G(n+1)\G(\hf + \b) \G(h-\frac{1}{2}+\b)} W_{h - 1/2 + n}\left(\b;\frac{1}{2},\frac{1}{2},h-\frac{1}{2},\frac{3}{2}-h\right)\\
&\qquad +\frac{d}{dn} M^{(K)}_n(\b|h).
\end{split}
\ee

The doubly split kernel is defined as
\be \label{kdsplitdefn}
K_\text{dsplit}(\a,\b|\D_\phi) = \frac{Q(\a)}{Q(-\b)} \int_0^1 \frac{dz}{z^2} \left(\frac{z}{1-z}\right)^{\D_\phi} k_{\a + 1/2}(z) k_{\b + 1/2}(1-z)\,,
\ee
so that
\be
K_\text{split} (\a,\b|\D_\phi) = K_\text{dsplit}(\a,\b|\D_\phi) + K_\text{dsplit}(-\a,\b|\D_\phi)\,.
\ee
Using the Mellin-Barnes representation of the hypergeometric functions we arrive at
\be
\begin{split}
&K_\text{dsplit}(\a,\b|\D_\phi) = N(\alpha) \frac{ 2 \alpha \beta }{\Gamma(\b + \hf)^2} \\
&\qquad \times \int[ds] \frac{\Gamma(s)^2 \Gamma(\b + \hf - s)^2 \G(\a - \hf + \D_\phi - s)\G(1 - \D_\phi + s)^2}{\G(\a + \frac{3}{2} - \D_\phi + s)}A^{\D_\phi -s,\D_\phi - \b - \hf}(\a)
\end{split}
\ee
with
\be
A^{p,q}(\a) \colonequals \frac{\G(\a + \frac{3}{2} - p)}{\G(1-p)^2 \G(\a + \hf)^2} \int [ds] \frac{\G(s)^2 \G(\a-s + \hf)^2 \G( 1- q - s)}{\G(\a + p - q -s + \hf)}\,.
\ee
In the limit where $\alpha$ is large and $p$ and $q$ are fixed we find that
\be
A^{p,q}(\a) = \a^{2q - 2p} \frac{\G(1-q)^2}{\G(1-p)^2} \left( 1 + O(\a^{-1})\right) \,.
\ee
Since $A^{p,q}(\a)$ has poles for negative real $\alpha$ and 
and therefore, in the large $\alpha$ and fixed $\beta$ limit 
\be
K_\text{dsplit} (\a,\b|\D_\phi) \sim \frac{2 N(\a)}{Q(-\b)} \G(\b - \D_\phi + 3/2)^2 (\a^2)^{\D_\phi - \b - 1}\,,
\ee
with $N(\a) = \hf Q(\a) Q(-\a)$. Notice also the obvious symmetry
\be
N(\b) K_\text{dsplit}(\a,\b|\D_\phi) = N(\a) K_\text{dsplit}(\b,\a|2-\D_\phi)\,,
\ee
and therefore in the large $\beta$ limit with fixed $\alpha$ we find
\be
K_\text{dsplit} (\a,\b|\D_\phi) \sim  Q(\a) \G(\a+ \D_\phi - 1/2)^2 (\b^2)^{1 - \D_\phi - \a}\,.
\ee

\section{\texorpdfstring{(Almost) recovering the identity in the $s$-channel}{(Almost) recovering the identity in the s-channel}}
\label{app:almostidentity}
The $s$-channel identity $f(z) = 1$ has a $t$-channel alpha space decomposition
\be
\fh_1^t(\b) = \frac{\G(\D_\phi - \hf \pm \b)\G(1-\D_\phi)}{\G(\hf \pm \b)\G(\D_\phi)}\,.
\ee
The poles at $\b = \D_\phi - \hf + n$ give a contribution
\be
\frac{\G(2\D + n -1)}{\G(\D)^2 n!}\,.
\ee
The idea in the main text is to use the doubly split kernel to get the $s$-channel density as a sum over $t$-channel blocks. Concretely this would mean
\be \label{fh1smaybe}
\fh_1^s(\a) \overset{?}{=} \sum_{n=0}^\infty \frac{\G(2\D + n -1)}{\G(\D)^2 n!} K_\text{dsplit}(\a,\D_\phi - 1/2 + n|\D_\phi) + (\a \leftrightarrow -\a)\,.
\ee
The summand behaves like $n^{-2 \a}$ and is therefore convergent for $\Re(\a) > 1/2$. We can conjecture what it is going to be by substituting the defining equation \eqref{kdsplitdefn} for the doubly split kernel and swapping the sum and the integral. This gives the $t$-channel block decomposition of the $s$-channel identity and therefore the sum should equal
\be
Q(\a) \int_0^1 \frac{dz}{z^2} k_{\a + 1/2}(z) = \frac{1}{2 \sin(\pi \a) \G(\frac{3}{2} \pm \a)}\,.
\ee
Numerical experiments for $\Re(\a) > 1/2$ confirm this conjecture and so we proceed by writing:
\be
\fh_1^s(\a) \overset{?}{=} \frac{1}{2 \sin(\pi \a) \G(\frac{3}{2} \pm \a)} + (\a \leftrightarrow -\a) \overset{!}{=} 0\,,
\ee
where the zero appears simply because the given function is odd in $\a$. This may seem contradictory but, as we discussed in the previous appendix, the alpha space density for the $s$-channel identity has support only for $\a = \pm 1/2$ and hence 0 is the correct answer almost everywhere! Clearly at $\a = 1/2$ various subtleties come into play: the integral in \eqref{kdsplitdefn} diverges but the overall factor $Q(\a)$ goes to zero, and furthermore the sum over $n$ ceases to converge. One may try to make sense of all this, but in our view it is more important to realize that the prescription in \eqref{fh1smaybe} requires an analytic continuation to compute the mirror image with $\a \to - \a$ and therefore it seems all but hopeless to recover any non-meromorphic pieces in the $s$-channel density. If one insists on getting the $s$-channel identity from the $t$-channel then one might be better off writing it as the limit of a function that has an analytic alpha space transform.

\section{List of integrals against blocks}
In this section we work out some integrals of simple functions against one-dimensional blocks $k_{\a + 1/2}(z)$. We first list a few integrals where $z$ ranges from 0 to 1. These play a role in the `split' discussions in the main text. None of the integrals here are particularly difficult and most have appeared in the literature before, see for example \cite{Hogervorst:2017sfd,Caron-Huot:2017vep,Cardona:2018dov,Liu:2018jhs,Sleight:2018ryu,Sleight:2018epi,Chen:2019gka}. We then discuss some integrals along the Hankel contour $C$ defined in section \ref{sec:moreanalytic}.

We first of all define
\be
\begin{split}
I^{(s)}_{p,q}(\a) &\colonequals \int_0^1 \frac{dz}{z^2} k_{\a + \hf}(z) z^p (1-z)^{-q}\\
&= \frac{\Gamma (1-q) \Gamma \left(p+\alpha -\frac{1}{2}\right)}{\Gamma \left(p-q+\alpha +\frac{1}{2}\right)} \, _3F_2\left(\begin{matrix}\alpha +\frac{1}{2},\alpha +\frac{1}{2},p+\alpha -\frac{1}{2}\\p-q+\alpha +\frac{1}{2},2 \alpha +1\end{matrix};1\right)\\
&= \frac{\Gamma (2 \alpha +1) \Gamma (1-q)^2 \Gamma \left(p+\alpha -\frac{1}{2}\right) \Gamma \left(-p+q+\alpha +\frac{1}{2}\right) }{\Gamma \left(\alpha +\frac{1}{2}\right)^2 \Gamma \left(-p+\alpha +\frac{3}{2}\right) \Gamma \left(p-q+\alpha +\frac{1}{2}\right)}
\, _3F_2\left(\begin{matrix}1-q,p-q,p-q\\p-q-\alpha +\frac{1}{2},p-q+\alpha +\frac{1}{2}\end{matrix};1\right)\\
& \phantom{=}+\frac{\Gamma (1-q) \Gamma \left(p+\alpha -\frac{1}{2}\right) \Gamma \left(p-q-\alpha -\frac{1}{2}\right)}{\Gamma (p-q)^2}
 \, _3F_2\left(\begin{matrix}\alpha +\frac{1}{2},\alpha +\frac{1}{2},-p+\alpha +\frac{3}{2}\\-p+q+\alpha +\frac{3}{2},2 \alpha +1\end{matrix};1\right)\,,
\end{split}
\ee
where the first ${}_3 F_2$ expression is defined for $q < 1$, where the integral converges near $z = 1$, and the second expression provides the analytic continuation to $q > 0$. Note the special cases
\be \label{usefulformula}
\begin{split}
I^{(s)}_{p,p}(\a) &\colonequals \int_0^1 \frac{dz}{z^2} k_{\a + \hf}(z) \left(\frac{z}{1-z}\right)^{p}\\
&= \frac{\Gamma (2 \alpha +1) \Gamma (1-p)^2 \Gamma \left(p+\alpha -\frac{1}{2}\right)}{\Gamma \left(\alpha +\frac{1}{2}\right)^2 \Gamma \left(-p+\alpha +\frac{3}{2}\right)}
\end{split}
\ee
and
\be
\begin{split}
I^{(s)}_{p,0}(\a) &\colonequals \int_0^1 \frac{dz}{z^2} k_{\a + \hf}(z) z^{p}\\
&= \frac{1}{p + \alpha -\frac{1}{2}}\, _3F_2\left(\begin{matrix}\alpha +\frac{1}{2},\alpha +\frac{1}{2},p+\alpha -\frac{1}{2}\\p+\alpha +\frac{1}{2},2 \alpha +1\end{matrix};1\right)\,.
\end{split}
\ee
For integrals of blocks against blocks we have, for two $s$-channel blocks,
\be
\begin{split}
&J^{(s)}(\a,\b) \colonequals \int_0^1 \frac{dz}{z^2} k_{\a + 1/2}(z) k_{\b + 1/2}(z) \\
&= \a \b Q(-\a)Q(-\b) \int [dt] \frac{\G(t + \b)\G(-t + \a) \G(\frac{1}{2} \pm t)^2}{\G(\b + 1 - t)\G(-\a + 1 + t)} \\
&= \a \b Q(-\a) Q(-\b) \Bigg\{ \frac{\G(\hf \pm \b)^2}{(\a + \b)\G(1 + \a - \b)\G(1 - \a + \b) } \\
&- \frac{d}{d\e} \left.\left[ \frac{\G(\hf + \a + \e)\G(\b -\hf -\e)}{\G(\hf + \a - \e)\G(\b + \frac{3}{2} + \e)}
\, _4F_3\left(\begin{matrix}\epsilon +1,\epsilon +1,-\alpha +\epsilon +\frac{1}{2},\alpha +\epsilon +\frac{1}{2}\\-\beta +\epsilon +\frac{3}{2},\beta +\epsilon +\frac{3}{2},2 \epsilon +1\end{matrix};1\right) \right]\right|_{\e = 0} \Bigg\}\,,
\end{split}
\ee
and for an $s$-channel against a $t$-channel block we have already defined the doubly split crossing kernel \eqref{kdsplitdefn}. For our purposes it is useful to introduce a slightly differently normalized function as follows:
\be
\begin{split}
&L^{(s)}_p(\a,\b) \colonequals \int_0^1 \frac{dz}{z^2} \left(\frac{z}{1-z}\right)^p k_{\a + 1/2}(z) k_{\b + 1/2}(1-z) = \frac{Q(-\b)}{Q(\a)} K_{\text{dsplit}}(\a,\b|p) = \\
&\frac{\Gamma (2 \beta +1) \Gamma \left(p+\alpha -\frac{1}{2}\right)^2 \Gamma (-p-\alpha +\beta +1)}{\Gamma \left(\beta +\frac{1}{2}\right)^2 \Gamma (p+\alpha +\beta )}
\, _4F_3\left(\begin{matrix}\alpha +\frac{1}{2},\alpha +\frac{1}{2},p+\alpha -\frac{1}{2},p+\alpha -\frac{1}{2}\\2 \alpha +1,p+\alpha -\beta ,p+\alpha +\beta\end{matrix};1\right)\\
&+ (\a,\b,p) \to (\b,\a, 2 - p)\,.
\end{split}
\ee

Next we consider the integrals along the Hankel contour $C$ that wraps the negative real axis. It is useful to note that
\be
\frac{1}{2\pi i}\int_C \frac{dz}{z^2}\left(\frac{z}{1-z}\right)^p = \frac{\sin(\pi(p-1))}{\pi(p - 1)}\,.
\ee
We have
\be
\begin{split}
I^{(C)}_{p,q}(\a) &\colonequals \frac{1}{2\pi i} \int_C \frac{dz}{z^2} k_{-\a + \hf}(z) z^p (1-z)^{-q}\\
&= \frac{1}{\pi} \sin(\pi(\a - p -1/2)) I^{(s)}_{p,p-q}(-\a)\\
&= \frac{\Gamma (-p+q+1)}{\Gamma \left(-p+\alpha +\frac{3}{2}\right) \Gamma \left(q-\alpha +\frac{1}{2}\right)}
 \, _3F_2\left(\begin{matrix}\frac{1}{2}-\alpha ,\frac{1}{2}-\alpha ,p-\alpha -\frac{1}{2}\\1-2 \alpha ,q-\alpha +\frac{1}{2}\end{matrix};1\right)\,,
\end{split}
\ee
which works for $p - q < 1$. It is worthwhile to also state the special case
\be
I^{(C)}_{p,0}(\a) = \frac{\Gamma (1-2 \alpha ) \Gamma (1-p)^2}{\Gamma \left(\frac{1}{2}-\alpha \right)^2 \Gamma \left(-p\pm\alpha +\frac{3}{2}\right)}\,.
\ee
For generic $p$ and $q$ we find poles whenever $\alpha$ is a positive integer which is when the conformal blocks themselves diverge. In the special case that $p$ is a half-integer we find that these poles disappear and we obtain an analytic function of $\alpha$. In either case the OPE coefficients are encoded in the density as described in the main text. Note that the double poles at integer values of $p$ for $q = 0$ originate from the divergence of the integral as $|z|\to \infty$.

Also,
\be
J^{(C)}(\a,\b) \colonequals \frac{1}{2\pi i} \int_C \frac{dz}{z^2} k_{-\a + 1/2}(z) k_{\b + 1/2}(z) = - \frac{1}{\pi} \sin(\pi(\a - \b)) J^{(s)}(-\a,\b)\,.
\ee
For generic $\beta$, the right-hand side again has kinematic poles when $\alpha$ is a positive integer. As discussed in the main text, this expression should be seen as providing an `analytic' continuation of the Kronecker delta.

Finally we have yet another crossing kernel. We define
\be
\begin{split}
K^{(C)}_p(\a,\b) &\colonequals \frac{1}{2\pi i} \int_C \frac{dz}{z^2} k_{-\a + 1/2}(z) k_{\b + 1/2}(1-z) \left(\frac{z}{1-z}\right)^p\\
&= \sum_{n,m = 0}^\infty \frac{(- \a + 1/2)_n^2(\b +1/2)_m^2}{(-2 \a + 1)_n (2\b +1)_m n! m!} \frac{\sin(\pi(p - \a - \b -1))}{\pi(p - \a - \b + n - m -1)}\,.
\end{split}
\ee
(In \texttt{Mathematica} it is faster to do one of the two infinite sums algebraically, which results in a ${}_3 F_2$ hypergeometric, before numerically evaluating the remaining sum. One can also just do the integral numerically.) One observes the symmetry
\be
K^{(C)}_p(\a,\b) = K^{(C)}_{2 - p}(-\b,-\a)\,,
\ee
which is obvious from the definition if one deforms the contour $C$ to lie parallel to the imaginary axis at $\Re(z) = 1/2$.

\section{Changing variables}
\label{app:alphatoJ}
In the main text we often encounter a density of the schematic form
\be
\frac{\tilde \lambda(\ab)}{\a - \tilde \b(\ab)}\,,
\ee
whereas we are more interested in a density of the schematic form
\be
\frac{\lambda(j)}{\a - \b(j)} + \text{regular}\,,
\ee
with $\ab = \a + j$. In order to match the location of the poles we must require that
\be
\b(j) - \tilde \b(j + \b(j)) = 0\,,
\ee
which defines $\b(j)$ in terms of $\tilde \b(\cdot)$. Secondly we have the equation
\be
\lambda(j) (\a - \tilde\b(j + \a)) = \tilde \lambda(\a + j) (\a - \b(j))  + (\a - \b(j)) (\a - \tilde \b(\a + j))\times \text{regular} \,.
\ee
Taking an $\alpha$ derivative and evaluating at $\alpha = \b(j)$ gives
\be
\lambda(j) ( 1 - \tilde \b'(j + \b(j)) ) = \tilde \lambda(\b(j) + j) \,,
\ee
and since
\be
\tilde \b'(j + \b(j)) = \frac{\b'(j)}{1 + \b'(j)}
\ee
we find
\be
\lambda(j) = (1 + \b'(j)) \tilde \lambda(\b(j) + j)\,,
\ee
which defines $\lambda(j)$ in terms of $\tilde \lambda(\cdot)$ and $\b(j)$.

\bibliographystyle{utphys}
\bibliography{References}

\providecommand{\href}[2]{#2}\begingroup\raggedright\begin{thebibliography}{10}

\bibitem{Hogervorst:2017sfd}
M.~Hogervorst and B.~C. van Rees, ``{Crossing symmetry in alpha space},''
  \href{http://dx.doi.org/10.1007/JHEP11(2017)193}{{\em JHEP} {\bfseries 11}
  (2017) 193},
\href{http://arxiv.org/abs/1702.08471}{{\ttfamily arXiv:1702.08471 [hep-th]}}.

\bibitem{Caron-Huot:2017vep}
S.~Caron-Huot, ``{Analyticity in Spin in Conformal Theories},''
  \href{http://dx.doi.org/10.1007/JHEP09(2017)078}{{\em JHEP} {\bfseries 09}
  (2017) 078},
\href{http://arxiv.org/abs/1703.00278}{{\ttfamily arXiv:1703.00278 [hep-th]}}.

\bibitem{flensted1973convolution}
M.~Flensted-Jensen and T.~Koornwinder, ``{The convolution structure for Jacobi
  function expansions},'' \href{http://dx.doi.org/10.1007/BF02388521}{{\em
  {Arkiv f{\"o}r Matematik}} {\bfseries 11} no.~1-2, (1973) 245--262}.

\bibitem{flensted1979jacobi}
M.~Flensted-Jensen and T.~H. Koornwinder, ``{Jacobi functions: The addition
  formula and the positivity of the dual convolution structure},''
  \href{http://dx.doi.org/10.1007/BF02385463}{{\em {Arkiv f{\"o}r Matematik}}
  {\bfseries 17} no.~1-2, (1979) 139--151}.

\bibitem{TK1}
T.~Koornwinder, \href{http://dx.doi.org/10.1007/978-94-010-9787-1}{``{Jacobi
  Functions and Analysis on Noncompact Semisimple Lie Groups},''} in {\em
  {Special Functions: Group Theoretical Aspects and Applications}}, pp.~1--85.
\newblock Springer, 1984.

\bibitem{Dobrev:1977qv}
V.~K. Dobrev, G.~Mack, V.~B. Petkova, S.~G. Petrova, and I.~T. Todorov,
  ``{Harmonic Analysis on the n-Dimensional Lorentz Group and Its Application
  to Conformal Quantum Field Theory},''
\href{http://dx.doi.org/10.1007/BFb0009678}{{\em Lect. Notes Phys.} {\bfseries
  63} (1977) 1--280}.

\bibitem{Dansthesis}
D.~Rutter, ``{Conformal Field Theory and the Alpha Space Transform \&
  Counterterms in Truncated Theories}.'' \url{http://etheses.dur.ac.uk/13106/},
  2019.
\newblock PhD thesis.

\bibitem{Isachenkov:2016gim}
M.~Isachenkov and V.~Schomerus, ``{Superintegrability of $d$-dimensional
  Conformal Blocks},''
  \href{http://dx.doi.org/10.1103/PhysRevLett.117.071602}{{\em Phys. Rev.
  Lett.} {\bfseries 117} no.~7, (2016) 071602},
\href{http://arxiv.org/abs/1602.01858}{{\ttfamily arXiv:1602.01858 [hep-th]}}.

\bibitem{Isachenkov:2017qgn}
M.~Isachenkov and V.~Schomerus, ``{Integrability of conformal blocks. Part I.
  Calogero-Sutherland scattering theory},''
  \href{http://dx.doi.org/10.1007/JHEP07(2018)180}{{\em JHEP} {\bfseries 07}
  (2018) 180},
\href{http://arxiv.org/abs/1711.06609}{{\ttfamily arXiv:1711.06609 [hep-th]}}.

\bibitem{Simmons-Duffin:2016wlq}
D.~Simmons-Duffin, ``{The Lightcone Bootstrap and the Spectrum of the 3d Ising
  CFT},'' \href{http://dx.doi.org/10.1007/JHEP03(2017)086}{{\em JHEP}
  {\bfseries 03} (2017) 086},
\href{http://arxiv.org/abs/1612.08471}{{\ttfamily arXiv:1612.08471 [hep-th]}}.

\bibitem{Komargodski:2012ek}
Z.~Komargodski and A.~Zhiboedov, ``{Convexity and Liberation at Large Spin},''
  \href{http://dx.doi.org/10.1007/JHEP11(2013)140}{{\em JHEP} {\bfseries 11}
  (2013) 140},
\href{http://arxiv.org/abs/1212.4103}{{\ttfamily arXiv:1212.4103 [hep-th]}}.

\bibitem{Fitzpatrick:2012yx}
A.~L. Fitzpatrick, J.~Kaplan, D.~Poland, and D.~Simmons-Duffin, ``{The Analytic
  Bootstrap and AdS Superhorizon Locality},''
  \href{http://dx.doi.org/10.1007/JHEP12(2013)004}{{\em JHEP} {\bfseries 12}
  (2013) 004},
\href{http://arxiv.org/abs/1212.3616}{{\ttfamily arXiv:1212.3616 [hep-th]}}.

\bibitem{Pappadopulo:2012jk}
D.~Pappadopulo, S.~Rychkov, J.~Espin, and R.~Rattazzi, ``{OPE Convergence in
  Conformal Field Theory},''
  \href{http://dx.doi.org/10.1103/PhysRevD.86.105043}{{\em Phys. Rev.}
  {\bfseries D86} (2012) 105043},
\href{http://arxiv.org/abs/1208.6449}{{\ttfamily arXiv:1208.6449 [hep-th]}}.

\bibitem{Mukhametzhanov:2018zja}
B.~Mukhametzhanov and A.~Zhiboedov, ``{Analytic Euclidean Bootstrap},''
  \href{http://dx.doi.org/10.1007/JHEP10(2019)270}{{\em JHEP} {\bfseries 10}
  (2019) 270},
\href{http://arxiv.org/abs/1808.03212}{{\ttfamily arXiv:1808.03212 [hep-th]}}.

\bibitem{Heemskerk:2009pn}
I.~Heemskerk, J.~Penedones, J.~Polchinski, and J.~Sully, ``{Holography from
  Conformal Field Theory},''
  \href{http://dx.doi.org/10.1088/1126-6708/2009/10/079}{{\em JHEP} {\bfseries
  10} (2009) 079},
\href{http://arxiv.org/abs/0907.0151}{{\ttfamily arXiv:0907.0151 [hep-th]}}.

\bibitem{Alday:2015eya}
L.~F. Alday, A.~Bissi, and T.~Lukowski, ``{Large spin systematics in CFT},''
  \href{http://dx.doi.org/10.1007/JHEP11(2015)101}{{\em JHEP} {\bfseries 11}
  (2015) 101},
\href{http://arxiv.org/abs/1502.07707}{{\ttfamily arXiv:1502.07707 [hep-th]}}.

\bibitem{Simmons-Duffin:2017nub}
D.~Simmons-Duffin, D.~Stanford, and E.~Witten, ``{A spacetime derivation of the
  Lorentzian OPE inversion formula},''
  \href{http://dx.doi.org/10.1007/JHEP07(2018)085}{{\em JHEP} {\bfseries 07}
  (2018) 085},
\href{http://arxiv.org/abs/1711.03816}{{\ttfamily arXiv:1711.03816 [hep-th]}}.

\bibitem{Kravchuk:2018htv}
P.~Kravchuk and D.~Simmons-Duffin, ``{Light-ray operators in conformal field
  theory},'' \href{http://dx.doi.org/10.1007/JHEP11(2018)102}{{\em JHEP}
  {\bfseries 11} (2018) 102}, \href{http://arxiv.org/abs/1805.00098}{{\ttfamily
  arXiv:1805.00098 [hep-th]}}.
[,236(2018)].

\bibitem{Alday:2015ewa}
L.~F. Alday and A.~Zhiboedov, ``{An Algebraic Approach to the Analytic
  Bootstrap},'' \href{http://dx.doi.org/10.1007/JHEP04(2017)157}{{\em JHEP}
  {\bfseries 04} (2017) 157},
\href{http://arxiv.org/abs/1510.08091}{{\ttfamily arXiv:1510.08091 [hep-th]}}.

\bibitem{Albayrak:2019gnz}
S.~Albayrak, D.~Meltzer, and D.~Poland, ``{More Analytic Bootstrap:
  Nonperturbative Effects and Fermions},''
  \href{http://dx.doi.org/10.1007/JHEP08(2019)040}{{\em JHEP} {\bfseries 08}
  (2019) 040},
\href{http://arxiv.org/abs/1904.00032}{{\ttfamily arXiv:1904.00032 [hep-th]}}.

\bibitem{Cornagliotto:2017snu}
M.~Cornagliotto, M.~Lemos, and P.~Liendo, ``{Bootstrapping the $(A_1,A_2)$
  Argyres-Douglas theory},''
  \href{http://dx.doi.org/10.1007/JHEP03(2018)033}{{\em JHEP} {\bfseries 03}
  (2018) 033},
\href{http://arxiv.org/abs/1711.00016}{{\ttfamily arXiv:1711.00016 [hep-th]}}.

\bibitem{Hogervorst:2016hal}
M.~Hogervorst, ``{Dimensional Reduction for Conformal Blocks},''
  \href{http://dx.doi.org/10.1007/JHEP09(2016)017}{{\em JHEP} {\bfseries 09}
  (2016) 017},
\href{http://arxiv.org/abs/1604.08913}{{\ttfamily arXiv:1604.08913 [hep-th]}}.

\bibitem{Cardona:2018dov}
C.~Cardona and K.~Sen, ``{Anomalous dimensions at finite conformal spin from
  OPE inversion},'' \href{http://dx.doi.org/10.1007/JHEP11(2018)052}{{\em JHEP}
  {\bfseries 11} (2018) 052},
\href{http://arxiv.org/abs/1806.10919}{{\ttfamily arXiv:1806.10919 [hep-th]}}.

\bibitem{Liu:2018jhs}
J.~Liu, E.~Perlmutter, V.~Rosenhaus, and D.~Simmons-Duffin, ``{$d$-dimensional
  SYK, AdS Loops, and $6j$ Symbols},''
  \href{http://dx.doi.org/10.1007/JHEP03(2019)052}{{\em JHEP} {\bfseries 03}
  (2019) 052},
\href{http://arxiv.org/abs/1808.00612}{{\ttfamily arXiv:1808.00612 [hep-th]}}.

\bibitem{Sleight:2018ryu}
C.~Sleight and M.~Taronna, ``{Anomalous Dimensions from Crossing Kernels},''
  \href{http://dx.doi.org/10.1007/JHEP11(2018)089}{{\em JHEP} {\bfseries 11}
  (2018) 089},
\href{http://arxiv.org/abs/1807.05941}{{\ttfamily arXiv:1807.05941 [hep-th]}}.

\bibitem{Sleight:2018epi}
C.~Sleight and M.~Taronna, ``{Spinning Mellin Bootstrap: Conformal Partial
  Waves, Crossing Kernels and Applications},''
  \href{http://dx.doi.org/10.1002/prop.201800038}{{\em Fortsch. Phys.}
  {\bfseries 66} no.~8-9, (2018) 1800038},
\href{http://arxiv.org/abs/1804.09334}{{\ttfamily arXiv:1804.09334 [hep-th]}}.

\bibitem{Chen:2019gka}
H.-Y. Chen and H.~Kyono, ``{On conformal blocks, crossing kernels and
  multi-variable hypergeometric functions},''
  \href{http://dx.doi.org/10.1007/JHEP10(2019)149}{{\em JHEP} {\bfseries 10}
  (2019) 149},
\href{http://arxiv.org/abs/1906.03135}{{\ttfamily arXiv:1906.03135 [hep-th]}}.

\end{thebibliography}\endgroup
\end{document}